\documentclass[onefignum, onetabnum]{onlineSIAM}

\usepackage{amssymb, amsfonts}
\usepackage{epstopdf}
\usepackage{bm}
\usepackage{cleveref}
\usepackage{graphicx, graphics, fancyhdr, subcaption}
\usepackage{scalerel}
\usepackage{hyperref}

\def\msquare{\mathord{\scalebox{0.32}[0.32]{\scalerel*{\Box}{\strut}}}}
\def\Msquare{\mathord{\scalebox{0.48}[0.48]{\scalerel*{\Box}{\strut}}}}

\usepackage{algorithmic}
\ifpdf
  \DeclareGraphicsExtensions{.eps,.pdf,.png,.jpg}
\else
  \DeclareGraphicsExtensions{.eps}
\fi

\usepackage{enumitem}
\setlist[enumerate]{leftmargin=.5in}
\setlist[itemize]{leftmargin=.5in}

\newsiamremark{remark}{Remark}
\newsiamremark{hypothesis}{Hypothesis}
\crefname{hypothesis}{Hypothesis}{Hypotheses}
\newsiamthm{claim}{Claim}

\headers{Simultaneous Affine Motion-compensated Image Reconstruction Technique}{Anh-Tuan Nguyen et al.}

\title{SAMCIRT: A Simultaneous Reconstruction and Affine Motion Compensation Technique for Four Dimensional Computed Tomography (4DCT)\thanks{Revision submitted on \today. This work is an extension of the conference papers \cite{Nguyen22_iCT, Nguyen22_SPIE, Nguyen23_ISBI}. Part of this study was conducted while the first author was affiliated with the imec-Vision Lab, Department of Physics, University of Antwerp, Universiteitsplein 1, Wilrijk 2610, Belgium.}

}

\author{Anh-Tuan Nguyen$^{\S, }$\thanks{Operations Research Lab (ORLab), Department of Computer Science, School of Computing, Phenikaa University, Nguyen Trac Street, Duong Noi Ward, Hanoi 12116, Vietnam.}
\and Jens Renders\thanks{imec-Vision Lab, University of Antwerp, Universiteitsplein 1, Wilrijk 2610, Belgium.}
\and Khoi-Nguyen Nguyen\thanks{Silverhead Innovation Strategies and Solutions SAS, 34 Avenue des Champs Elysées, 75008 Paris, France.}
\and Tat-Dat To\thanks{Institut de Mathématiques de Jussieu – Paris Rive Gauche, 4 Place Jussieu, 75252 Paris, France.\newline
Corresponding email: \email{tuan.nguyenanh2@phenikaa-uni.edu.vn}{ (Anh-Tuan Nguyen).}}
\and Domenico Iuso\footnotemark[3]
\and Yves Maris\footnotemark[3] 
}

\makeatletter
\newcommand*{\addFileDependency}[1]{
  \typeout{(#1)}
  \@addtofilelist{#1}
  \IfFileExists{#1}{}{\typeout{No file #1.}}
}
\makeatother

\ifpdf
\hypersetup{
  pdftitle={SMART},
  pdfauthor={Anh-Tuan Nguyen et al.}
}
\fi

\begin{document}

\maketitle

\begin{abstract}
In four-dimensional computed tomography (4DCT), 3D images of moving or deforming samples are reconstructed from a set of 2D projection images. Recent techniques for iterative motion-compensated reconstruction either necessitate a reference acquisition or alternate image reconstruction and motion estimation steps. In these methods, the motion estimation step involves the estimation of either complete deformation vector fields (DVFs) or a limited set of parameters corresponding to the affine motion, including rigid motion or scaling. The majority of these approaches not only rely on nested iterations, thereby increasing computational complexity and constraining potential acceleration, but also fail to provide a theoretical proof of convergence for their proposed iterative schemes. On the other hand, the latest MATLAB and Python image processing toolboxes lack the implementation of analytic adjoints of affine motion operators for 3D object volumes, which does not allow gradient methods using exact derivatives towards affine motion parameters. In this work, we propose the Simultaneous Affine Motion-Compensated Image Reconstruction Technique (SAMCIRT)- an efficient iterative reconstruction scheme that combines image reconstruction and affine motion estimation in a single update step, based on the analytic adjoints of the motion operators then exact partial derivatives with respect to both the reconstruction and the affine motion parameters. Moreover, we prove the separated Lipschitz continuity of the objective function and its associated functions, including the gradient, which supports the convergence of our proposed iterative scheme, despite the non-convexity of the objective function with respect to the affine motion parameters. Results from simulation and real experiments show that our method outperforms the state-of-the-art CT reconstruction with affine motion correction methods in computational feasibility and projection distance. In particular, this allows accurate reconstruction for a real, nonstationary diamond, showing a novel application of 4DCT.
\end{abstract}

\begin{keywords}
Iterative tomographic reconstruction, affine motion models, simultaneous motion estimation.
\end{keywords}

\begin{AMS}
65K10, 68U10, 68W01, 92C55, 94A08.
\end{AMS}

\section{Introduction}
Four-dimensional computed tomography (4DCT) is a crucial aspect of the broader field of CT imaging that captures 3D objects in motion or the changes of the microstructure, or in a more general term, in the evolution of their shape over time. 4DCT reconstruction techniques restore 4D images that can be regarded as changing 3D volumes over time. Reconstruction is based on one or multiple sets $360^{\circ}$ of 2D projection images, and applications of 4DCT are widely found in medicine along with materials science, with the expectation of improving the resolution in both the spatial and temporal domains \cite{Taillandier-Thomas16}. 

However, object motion during the 4DCT scan introduces motion artifacts in the reconstructed images if it is not taken into account in the reconstruction process. In medical applications, these artifacts complicate both diagnosis and treatment. To address this issue, motion-compensated reconstruction techniques have been proposed to reduce these artifacts and allow accurate 3D volume reconstruction\cite{Korreman12}. 


4D tomographic reconstruction methods in the history can be categorized into two different groups. The first group of methods incorporate the information on both the spatial and temporal domains without parameterizing the motion. A typical example is \cite{Kadu23Single} that incorporates an optical flow model in the dynamic computerized tomography framework when a time component is considered. On the other hand, \cite{Chung18} proposes an efficient method based on Krylov subspaces. Moreover, \cite{Pasha2023Computational} proposes a general computationally efficient set of methods for spatiotemporal dynamic inverse problems with total variation and group sparsity priors. More recently, in \cite{lan2023Spatiotemporal}, a Bayesian approach was proposed based on Besov priors to preserve the edge and promote the sparsity. Some of the other references that provide test problems and methods to solve dynamic problems are presented in \cite{Jorgensen21Core, Papoutsellis2021Core, Pasha2024Trips}. All these references show ways to solve dynamic problems and propose several ways to incorporate temporal information into the regularization term. Nonetheless, these methods do not estimate motions. As a result, it is challenging to quantify the change in the object during the CT scan, regardless of whether the change is local or global. The other group of methods aim to estimate the motion, most published recently, and is split into two subgroups. The first subgroup consists of methods that estimate general-affine motion parameters directly from projection data (e.g., \cite{VanNieuwenhove17_TIP}), whereas the methods in the second subgroup alternate image reconstruction and motion estimation of full deformation vector fields (DVFs) or a limited set of affine motion parameters (e.g., \cite{Chen19, Chung17, Lucka18, Zang19, Zehni20, Renders23ImWIP, Nguyen23_ISBI}). However, these methods are either not validated on real data or are validated but the optimization strategies rely on the optimizers that require parameter tuning and nested iterations, which increases the computational complexity and reduce the potential acceleration of these algorithms. In addition, no theoretical proof of convergence is given for the proposed iterations. Furthermore, it is worth noticing that all the methods lack proper study in joint image reconstruction and general affine motion estimation with analytical gradients of the motion towards both the reconstruction and the general affine motion parameters for 3D object volumes, which are not yet implemented in the latest MATLAB and Python image processing toolboxes. Although several continuous time-step motion models of simulated objects and discrete-time motion models with a real object are considered as in our previous studies \cite{Nguyen22_iCT, Nguyen22_SPIE, Nguyen23_ISBI}, these motion models have not been applied yet to real objects with possibly compatible motion models. Moreover, substantial efforts are still required to establish the mathematical correctness of the proposed method. Furthermore, quantitative comparisons with state-of-the-art dynamic CT reconstruction methods are currently missing. 

 Consequently, in this paper, we present an efficient iterative reconstruction and motion correction technique for 4DCT based on the exact adjoints and gradients of the motion, which can incorporate any parameterizable DVF including general affine motions. Motion models can be deduced either from the characterization of the material of which the object is made, or from the observation of its dynamic behavior during scanning. Furthermore, we rely on an objective function that depends on both the object volume and the parameterized DVFs, whose partial derivatives with respect to all these variables are analytically described. This allows simultaneous reconstruction and general affine motion estimation using gradient-based optimization strategies. More importantly, we present a theoretical foundation for the convergence of our proposed iterative scheme. On the other hand, we extend our findings to real data, followed by quantitative comparisons of the results obtained by our method with the results obtained by the state-of-the-art reconstruction and affine motion estimation methods. In particular, our method allows micro-scale structures of a real diamond whose movement is disruptive for the analysis when using conventional 3DCT reconstruction techniques, which leads to a novel application in 4DCT.

The structure of the rest of the paper is organized as follows: in \cref{sec:method}, we present a mathematical formulation for the 4DCT forward model, in which the CT system operator is inspired by \cite{VanEyndhoven12, Zehni20} and can be decomposed as the multiplication of the CT projector and the global affine motion operator. This is then followed by our underlying algorithm that allows accurate reconstruction and affine motion parameter estimation simultaneously, by considering the corresponding iterative schemes. More specifically, this is a gradient method whose partial derivatives are given analytically. A crucial component of this section is the analysis of the compactness of the reconstruction and motion domains, together with the establishment of separated Lipschitz continuity for the objective function and its associated mappings, including the gradient. These properties provide a theoretical foundation for the convergence of the proposed iterative scheme, despite the non-convexity of the objective function. In \cref{sec:datasets}, we describe simulated and real projection datasets relevant to the considered motion models, which are used in the validation process in \cref{sec:validation}, wherein the comparison with several state-of-the-art methods is given. Another crucial contribution is mentioned in this section is the validation of our method on a real diamond scan, regarded as the very first study in 4DCT that supports diamond inspection. In \cref{sec:discussion}, we give an overall discussion on the method, followed by a conclusion of the study in \cref{sec:conclusion}. Finally, \cref{sec:futurework} provides a future outlook.

\section{Proposed method}\label{sec:method}
This section presents the mathematical formulation of affine motions used in dynamic CT processes for modeling parameterizable DVFs, which allows accurate reconstruction and estimation of the corresponding affine motion parameters. Moreover, a numerical standard is introduced to quantitatively compare the reconstruction and motion estimation results with the results from the state-of-the-art reconstruction and affine motion estimation methods.

\subsection{Forward model}\begin{definition}{(Radon transform).}
For a compactly supported 3D object function $\bm{f}(\bm{u})$ with $\bm{u} \in \mathbb{R}^3$, the Radon transform $\mathcal{R}$ of $\bm{f}$ over the plane orthogonal to the unit normal vector $
\bm{n} = \begin{bmatrix}\sin \theta \cos \phi, \; \sin \theta \sin \phi, \; \cos \theta \end{bmatrix}^{\top}$ at signed distance $d$ from the origin is defined as
\begin{equation}
    \mathcal{R}\{ \bm{f} \}(d, \bm{n}) := \int_{\mathbb{R}^3} \bm{f}(\bm{u}) \, \delta\!\left(d - \langle \bm{n}, \bm{u} \rangle \right) \, d\bm{u},
\end{equation}
where $\delta$ is the Dirac delta distribution and $\langle \cdot, \cdot \rangle$ is the standard inner product.
\end{definition}

A 4D object can be represented as a sequence of $n$ 3D volumes $\bm{f}_1, \dots, \bm{f}_n$, each corresponding to the object at a given point in time. Each volume $\bm{f}_i$ is discretized and vectorized into $\bm{x}_i \in \mathbb{R}^{n_\text{vol}}$, where $n_\text{vol}$ is the number of voxels of the 3D grid. 

The acquisition process can be modeled as a collection of \emph{subscans}, where the object is assumed static during each subscan. A subscan consists of one or more consecutive projections, and we denote by $\bm{\tilde{W}}_i$ the corresponding nonlinear CT projector for the $i^\text{th}$ subscan. The measurement model is then given by the system of nonlinear equations
\begin{equation}\label{eq:phys_proc}
    \bm{\tilde{W}}_i(\bm{f}_i) = \bm{b}_i, \quad i = 1,\dots,n,
\end{equation}
where $\bm{b}_i \in \mathbb{R}^{n_i \cdot n_\text{det}}$ is the acquired projection data, $n_i$ is the number of projections in the subscan, and $n_\text{det}$ is the number of detector elements.

For tractability, and as is common in algorithmic studies, we adopt a simplified linearized model that neglects physical effects such as scatter and the polyenergetic nature of the X-ray spectrum (e.g., beam hardening). Under this approximation, the nonlinear systems \eqref{eq:phys_proc} are reduced to the linear systems
\begin{equation}\label{eq:math_proc}
    \bm{W}_i \bm{x}_i = \bm{b}_i, \quad i = 1,\dots,n,
\end{equation}
where $\bm{W}_i \in \mathbb{R}^{(n_i \cdot n_\text{det}) \times n_\text{vol}}$ is the projection matrix that maps the volume $\bm{x}_i$ to the corresponding projection data $\bm{b}_i$.

These may be represented as the following system:
\begin{equation}
\label{eq:simple4d}
\begin{bmatrix}
    \bm{W}_1 &   0      & 0      & 0 \\
      0      & \bm{W}_2 & 0      & 0 \\
      0      &   0      & \ddots & 0\\
      0      &   0      &     0  & \bm{W}_n
\end{bmatrix}
\begin{bmatrix}
    \bm{x}_1 \\
    \bm{x}_2 \\
    \vdots \\
    \bm{x}_n
\end{bmatrix}
=
\begin{bmatrix}
    \bm{b}_1 \\
    \bm{b}_2 \\
    \vdots \\
    \bm{b}_n
\end{bmatrix}.
\end{equation}

Assume that the unknown image $\bm{x}_i$ of the $i^\text{th}$ subscan is a transformed version of the unknown static object volume $\bm{x}$ under an affine motion model $\mathcal{M}$ depends on the parameter $\bm{p}_i \in \mathbb{R}^{12}$, i.e.,
\begin{equation}\label{moving_image}
    \bm{x}_i = \mathcal{M}\left(\bm{p}_i \right) \bm{x}.
\end{equation}

Here, $\mathcal{M}\left(\bm{p}_i \right) \in \mathbb{R}^{n_\text{vol} \times n_\text{vol}}$ is a warping operator that models the motion as an affine transformation, and the unknown object $\bm{x}$ is assumed to be static in the first projections.

\subsection{Analytic derivation of general affine motion operator with respect to affine motion parameters}
Let us consider the following affine transformation:
\begin{equation}    
{\bm{f}_i}(\bm{u}) = \bm{f}(\bm{A}_i\bm{u}+\bm{t}_i),
\end{equation}
where the affine matrix $\bm{A}_i \in \mathbb{R}^{3 \times 3}$ and the three elements of the translation $\bm{t}_i \in \mathbb{R}^3$. Here, the voxel coordinate $\bm{u}$ of the original 3D object $\bm{f}$ is shifted into the 3D warped object ${\bm{f}_i}$ under a differentiable affine map with respect to the affine motion parameter $\bm{p}_i$ as a collection of the elements of $\bm{A}_i$ and $\bm{t}_i$. Moreover, this change of coordinates is generalized as the following:
\begin{equation}
    \bm{\widehat{\bm{u}}} = \bm{A}_i \bm{u} + \bm{c} - \bm{A}_i\bm{c} + \bm{t}_i,
\end{equation}
where $\bm{c} \in \mathbb{R}^3$ is the center of the system of coordinates. 

As affine transformations are differentiable, the motion operator $\mathcal{M}$ then becomes differentiable if missing voxels in the warped images are interpolated by differentiable maps (e.g., linear, quadratic or cubic). This approach provides an analytic derivative $\nabla \mathcal{M}(\bm{p}_i)\bm{x}$ with respect to the affine motion parameter $\bm{p}_i$ for the warped image $\mathcal{M}(\bm{p}_i)\bm{x}$, aside an adjoint operator $\mathcal{M}\left(\bm{p}_i \right)^{\top}$ \cite{Renders21}, where $\bm{x}$ represents the set of all voxels in the 3D image representing the object $\bm{f}$.

\subsection{Dynamic CT model}
By decomposing the full stack of transformed images by the relation \cref{moving_image} and substituting the result into \cref{eq:simple4d}, it yields:
\begin{equation}
\label{eq:motion4d}
\begin{bmatrix}
    \bm{W}_1 &   0 & 0 & 0 \\
      0 & \bm{W}_2 & 0 & 0 \\
      0 &   0 & \ddots & 0\\
      0 &   0 &     0 & \bm{W}_n
\end{bmatrix}
\begin{bmatrix}
     \mathcal{M}\left(\bm{p}_1\right)  \\
     \mathcal{M}\left(\bm{p}_2\right) \\
    \vdots \\
     \mathcal{M}\left(\bm{p}_n \right)
\end{bmatrix}
\bm{x}
=
\begin{bmatrix}
    \bm{b}_1 \\
    \bm{b}_2 \\
    \vdots \\
    \bm{b}_n
\end{bmatrix},
\end{equation}
 or more concisely:
\begin{equation}
\label{eq:motion4dshort}
    \bm{W}\bm{\mathcal{M}}(\bm{p})\bm{x}=\bm{b},
\end{equation}
where
\begin{equation}
    \bm{W} =
    \begin{bmatrix}
        \bm{W}_1 & 0 & 0 & 0 \\
        0 & \bm{W}_2 & 0 & 0 \\
        0 & 0 & \ddots & 0 \\
        0 & 0 & 0 & \bm{W}_n
    \end{bmatrix} \in \mathbb{R}^{\left(\sum_{i=1}^{n} n_i \cdot n_{\text{det}} \right) \times \left(n \cdot n_\text{vol} \right)},
\end{equation}

\begin{equation}
    \bm{p} =
    \begin{bmatrix}
        \bm{p}_1 \\
        \bm{p}_2 \\
        \vdots \\
        \bm{p}_n
    \end{bmatrix} \in \mathbb{R}^{12n},
    \bm{\mathcal{M}}(\bm{p}) =
    \begin{bmatrix}
        \mathcal{M}(\bm{p}_1) \\
        \mathcal{M}(\bm{p}_2)\\
        \vdots  \\
        \mathcal{M}(\bm{p}_n)
    \end{bmatrix} \in \mathbb{R}^{\left(n \cdot n_\text{vol} \right) \times n_\text{vol}}, \text{ and }
    \bm{b} =
    \begin{bmatrix}
        \bm{b}_1 \\
        \bm{b}_2 \\
        \vdots \\
        \bm{b}_n
    \end{bmatrix} \in \mathbb{R}^{\sum_{i=1}^n n_i \cdot n_\text{det}}.
\end{equation}

\subsection{Optimization strategies}
In order to solve the nonlinear system \cref{eq:motion4dshort} in least-square sense, a modified version of the gradient method in \cite{VanEyndhoven12, Zehni20} that considers a joint estimation of the motion parameters $\bm{p}$ along with the reconstruction image $\bm{x}$, is used to solve the following minimization problem:
\begin{equation}
    \left[ \widehat{\bm{x}}, \widehat{\bm{p}} \right] = \text{argmin}_{\bm{x}, \bm{p}}  g\left(\bm{x}, \bm{p} \right),
\end{equation}
where
\begin{equation}\label{eq:formal_obj}
    g\left(\bm{x}, \bm{p} \right) = \frac{1}{2} \left  \| \bm{W}\bm{\mathcal{M}}(\bm{p})\bm{x} - \bm{b} \right \|^2,
\end{equation}
with $\left(\bm{x}, \bm{p} \right) $ is in a compact extended domain $\mathcal{D} = \mathcal{X} \times \mathcal{P} \subset \mathbb{R}^{n_\text{vol}} \times \mathbb{R}^{12n}$ (explanation of the compactness is in \cref{convergenceproof}) and $\left \| \cdot \right \|$ is the usual Euclidean norm.

Our problem can be viewed as a separable nonlinear least-square problem, for which there exist algorithms that may converge faster with convergence results \cite{GP73, GP03, OR13, EJ24, vLA16}. However, we avoid these techniques as the objective function may not be always differentiable with respect to the motion parameters, which cannot provide exact partial derivatives used for the steps of gradient methods.
 
As an ill-posed problem, the objective function \cref{eq:formal_obj} should include regularization terms for both the reconstruction and the motion parameters. However, if the global motion operator $\bm{\mathcal{M}}$ is sufficiently smooth, it is worthwhile to investigate whether accurate reconstruction and motion parameters can be obtained without regularization. In practice, TV-$l_1$ is commonly used as a smoothing regularization in parameterizable optical flows, implemented with the exact adjoints of the motion operators \cite{Renders21}. The gradient of this objective function is analytically given by $\nabla g = \left[ \left [ \nabla_{\bm{x}} g \right ]^{\top}, \left [ \nabla_{\bm{p}} g \right ]^{\top}  \right ]^{\top}$, with
\begin{equation}\label{eq:f_x}
    \nabla_{\bm{x}} g \left(\bm{x}, \bm{p} \right) = \bm{\mathcal{M}} \left(\bm{p}  \right)^{\top} \bm{W}^{\top}\bm{r},
\end{equation}
and
\begin{equation}\label{eq:f_p}
    \nabla_{\bm{p}} g(\bm{x}, \bm{p} ) = \left[\nabla\bm{\mathcal{M}} \left(\bm{p} \right)\bm{x} \right]^{\top} \bm{W}^{\top}\bm{r},
\end{equation}
where $\bm{r} = \bm{W}\bm{\mathcal{M}}\left(\bm{p}  \right)\bm{x}- \bm{b}$ is the residue of the dynamic CT system.

The implementation of the global motion operator $\bm{\mathcal{M}} \left(\bm{p} \right)$, its adjoint $\bm{\mathcal{M}} \left(\bm{p} \right)^{\top}$ and its derivative $\nabla\bm{\mathcal{M}}\left(\bm{p} \right)$ are all done beforehand by the ImWIP \cite{Renders23ImWIP}- an open-source warping toolbox that contains matrix-free and GPU-accelerated implementations of cubic image warping, its adjoint and its analytical derivatives. The projector $\bm{W}$ and its adjoint $\bm{W}^{\top}$ are provided by the ASTRA Toolbox \cite{VanAarle16}.

In theory, an ordinary update step using a gradient method would require a single step size $\gamma^k$ to be chosen for the update of both the image and the motion parameters:
\begin{equation}
    \left[\bm{x}^{k+1}, \bm{p}^{k+1} \right] = \left[\bm{x}^{k}, \bm{p}^{k} \right] - \gamma^k \nabla g\left(\bm{x}^k, \bm{p}^k \right).
\end{equation}

However, to accommodate for the fact that the motion parameters play a fundamentally different role in the system, and that these influence the objective function with different sensitivities, the update of the gradient method is split in their corresponding parts with an independent stepsize for each part:
\begin{equation}\label{eq:GDBB_x}
    \bm{x}^{k+1} = \bm{x}^{k} - \gamma_{\bm{x}}^k \nabla_{\bm{x}} g\left(\bm{x}^k, \bm{p}^{k} \right),
\end{equation}
and
\begin{equation}\label{eq:GDBB_p}
    \bm{p}^{k+1} = \bm{p}^{k} - \gamma_{\bm{p}}^k \nabla_{\bm{p}} g\left(\bm{x}^k, \bm{p}^{k} \right).
\end{equation}

Because of the large scale of the 3D object volume sizes, it is challenging to directly implement the formulas \cref{eq:f_x} and \cref{eq:f_p}. Nonetheless, the objective function \cref{eq:formal_obj} can be rewritten as  $g\left(\bm{x}, \bm{p} \right) = \frac{1}{2}\sum\limits_{i=1}^{n} \left \| \bm{W}_i \mathcal{M}\left( \bm{p}_i\right) \bm{x} - \bm{b}_i\right \|^2$ in the sparsity of the projection operator $\bm{W}$. 

The partial derivative of the proposed objective function with respect to the reconstruction \cref{eq:f_x} is then given by:
\begin{equation}\label{eq:f_x_simplified}
    \nabla_{\bm{x}} g\left(\bm{x}, \bm{p} \right) = \sum\limits_{i=1}^{n} {\mathcal{M}} \left(\bm{p}_i  \right)^{\top} \bm{W}_i^{\top} \left[ \bm{W}_i \mathcal{M}\left(\bm{p}_i \right) \bm{x} - \bm{b}_i \right].
\end{equation}

Similarly, the partial derivative of it with respect to each parameter based on the general form as given by the equation \cref{eq:f_p}, is given as follows:
\begin{equation}\label{eq:f_pi_simplified}
    \nabla_{\bm{p}_i} g(\bm{x}, \bm{p} ) = \left[\nabla {\mathcal{M}} \left(\bm{p}_i \right)\bm{x} \right]^{\top} \bm{W}_i^{\top} \left[ \bm{W}_i \mathcal{M}\left(\bm{p}_i \right) \bm{x} - \bm{b}_i \right] \quad \forall i= \overline{1, n}, 
\end{equation}
which remains invariant with respect to the motion parameters associated with the other subscans. Importantly, allow us to note that the objective function $g$ is \textit{non-convex} with respect to the motion parameters $\bm{p}$. 

\subsection{The compactness of the domains, the Lipschitz continuity of the relevant functions and the convergence of the schemes \eqref{eq:GDBB_x} and \eqref{eq:GDBB_p}}\label{convergenceproof}

In the scenarios considered in our study, the scanned objects are always placed within the field of view (FOV) of the CT scanner. As a result, it is natural to acknowledge that these are bounded within a closed area and that their motion occurs entirely within this area. In other words, the extended domain $\mathcal{P}$ of the smooth affine global motion operator ${\bm{\mathcal{M}}}\left(\cdot \right)$ is compact. Also, from a theoretical perspective, $\left \| \bm{x} \right \|$ may be very large if the voxels within the FOV have high values. However, physics imposes limitations: the attenuation coefficient has an upper bound determined by the properties of materials and it cannot exceed the attenuation coefficient of heavy metals at a given X-ray energy \cite{Buzug2008}. Hence, the reconstruction domain $\mathcal{X}$ is also compact. Without loss of generality, let us assume a stronger property that the global motion operator $\bm{\mathcal{M}} \in \mathcal{C}^{2}\left(\mathcal{P} \right)$. Hence, there exists an upper bound $B_{\text{vol}} < +\infty$ for the object volume, i.e.,
\begin{equation}    
\sup_{\bm{x} \in \mathcal{X}} \left  \|\bm{x} \right \| \le B_{\text{vol}};
\end{equation}
and upper bounds $B_{\bm{\mathcal{M}}}, B_{\nabla{\bm{\mathcal{M}}}}, B_{\nabla {\bm{\mathcal{M} \bm{x}}}}  \text{ and } B_{\nabla ^2{\bm{\mathcal{M} \bm{x}}}}  < +\infty$ for the global operators, i.e.,
\begin{align}    
    \sup_{\bm{p} \in \mathcal{P}} \left \| {\bm{\mathcal{M}}}(\bm{p}) \right \| \le B_{\bm{\mathcal{M}}} \text{, }  &\sup_{\bm{p} \in \mathcal{P}}  \left\| {\nabla {\bm{\mathcal{M}}}}(\bm{p}) \right\| \le B_{\nabla{\bm{\mathcal{M}}}}, \\
  \sup_{\bm{p} \in \mathcal{P}, \bm{x} \in \mathcal{X}}  \left\| {\nabla {\bm{\mathcal{M}}}}(\bm{p})\bm{x} \right\| \le B_{\nabla{\bm{\mathcal{M}}}\bm{x}} &\text{, and }  \sup_{\bm{p} \in \mathcal{P}, \bm{x} \in \mathcal{X}}  \left\| {\nabla^2 {\bm{\mathcal{M}}}(\bm{p}) \bm{x}} \right\| \le B_{\nabla^2{\bm{\mathcal{M}}\bm{x}}}.
\end{align}

In our study, CT scans are modeled as closed physical systems, which implies the boundedness of the norms of the relevant variables and operators. This boundedness constitutes the essential property underlying the proof of the Lipschitz continuity of the proposed objective function $g$ and its gradient $\nabla g$. Our proof is sketched as follows: First of all, we verify the separated Lipschitz continuity of the projection function $\bm{b} \left(\bm{x}, \bm{p} \right) = \bm{W}\bm{\mathcal{M}}\left(\bm{p} \right)\bm{x}$, the norm $ \left \| \bm{b} \left(\bm{x}, \bm{p} \right) \right \|$, the Jacobian  $\mathcal{J}_{\bm{b}} \left(\bm{x}, \bm{p} \right)$ and the objective function $g \left(\bm{x}, \bm{p} \right)$. This property of these functions are regarded as lemmas for further explanation in the proof of the same property of the gradient function $ \nabla \left \| {\bm{b}} \left(\bm{x}, \bm{p} \right) \right \|$, and finally so is the gradient function $\nabla g \left(\bm{x}, \bm{p} \right)$.

\begin{proposition}\label{prop:Mpx}
The projection function\footnote{Not be confused with the projection data $\bm{b}$.} $ \bm{b}\left(\bm{x}, \bm{p} \right) = \bm{W}\bm{\mathcal{M}\left(\bm{p}\right)}\bm{x}$ is separated Lipschitz continuous in $\bm{x}$ and $\bm{p}$, then so are the norm function $\left \| \bm{b}\left(\bm{x}, \bm{p} \right) \right \|$, the Jacobian $\mathcal{J}_{\bm{b}} \left(\bm{x}, \bm{p} \right)$, and the objective function $g\left(\bm{x}, \bm{p} \right)$. 
\end{proposition}

\begin{proof} 
First of all, we acknowledge that in reality the projection function is not vanished everywhere excepts at the image $\bm{x} \equiv \bm{0}$ of the null object. For any $\bm{x} \in \mathcal{X}$, $\bm{y} \in \mathcal{X}$, $\bm{p} \in \mathcal{P}$, and $\bm{q} \in \mathcal{P}$, we break our proof into different parts.\\

\textit{Part 1: Separated Lipschitz continuity of $\bm{b}\left(\bm{x}, \bm{p}\right)$ and $\left \| \bm{b}\left(\bm{x}, \bm{p}\right) \right \|$.}\label{part:Lipschitz_b_normb}

Indeed, for the reconstruction variable we have:  
\begin{align}\label{eqn:bxp_byp}
    \left \|\bm{b}\left(\bm{x}, \bm{p} \right) - \bm{b}\left(\bm{y}, \bm{p} \right) \right \| &= \left \| \bm{W} \bm{\mathcal{M}}\left(\bm{p}\right)\bm{x} -  \bm{W} \bm{\mathcal{M}}\left(\bm{p}\right)\bm{y}\right \| = \left \| \bm{W} \bm{\mathcal{M}}\left(\bm{p}\right) \left[\bm{x} -\bm{y} \right] \right \| \\ \nonumber
    & \le \left\|  \bm{W} \right \| \cdot \left \| \bm{\mathcal{M}}\left(\bm{p}\right) \right\| \cdot \left \|\bm{x} -\bm{y} \right \| \le \left \|\bm{W} \right \| \cdot B_{\bm{\mathcal{M}}} \cdot \left \|\bm{x} -\bm{y} \right \|.
\end{align}
Also, the same for the motion parameters:
\begin{equation}
    \left \|\bm{b}\left(\bm{x}, \bm{p} \right) - \bm{b}\left(\bm{x}, \bm{q} \right) \right \| = \left \| \bm{W} \bm{\mathcal{M}}\left(\bm{p}\right)\bm{x} -  \bm{W} \bm{\mathcal{M}}\left(\bm{q}\right)\bm{x}\right \| \le \left \| \bm{W} \right \| \cdot \left \| \bm{\mathcal{M}}\left(\bm{p}\right) \bm{x} - \bm{\mathcal{M}}\left(\bm{q}\right) \bm{x} \right \|. 
\end{equation}

Let us recall the intermediate value theorem in Banach spaces:
\begin{equation}
 \left \| {\bm{\mathcal{M}}} (\bm{p})\bm{x}  - {\bm{\mathcal{M}}}(\bm{q})\bm{x}  \right \|  \le {B_{\nabla{\bm{\mathcal{M}}}\bm{x} }} \cdot \left \| \bm{p} - \bm{q} \right \|.   
\end{equation}

As a result,
\begin{equation}\label{eqn:bxp_bxq}
    \left \|\bm{b}\left(\bm{x}, \bm{p} \right) - \bm{b}\left(\bm{x}, \bm{q} \right) \right \| \le \left\|  \bm{W} \right \| \cdot  {B_{\nabla{\bm{\mathcal{M}}}\bm{x}}} \cdot \left \| \bm{p} - \bm{q} \right \|.  
\end{equation}

Consequently, the projection function $ \bm{b}\left(\bm{x}, \bm{p} \right) =\bm{W}\bm{\mathcal{M}\left(\bm{p}\right)}\bm{x}$ is separated Lipschitz continuous in $\bm{x}$ and $\bm{p}$ with the corresponding coefficients $L_{\bm{x}}^{\bm{b}} =  \left \| \bm{W} \right \| \cdot B_{\bm{\mathcal{M}}}$ and $L_{\bm{p}}^{\bm{b}} = \left \| \bm{W} \right \| \cdot {B_{\nabla{\bm{\mathcal{M}}}\bm{x}}} $, respectively. Moreover, the following reverse triangle inequalities hold:
\begin{equation}\label{eqn:abs_bxp_byp}
    \left | \left\| \bm{b}\left(\bm{x}, \bm{p} \right) \right \| - \left\| \bm{b}\left(\bm{y}, \bm{p} \right) \right \| \right | \le \left\| \bm{b}\left(\bm{x}, \bm{p} \right) -  \bm{b}\left(\bm{y}, \bm{p} \right) \right \|
\end{equation}
and 
\begin{equation}\label{eqn:abs_bxp_bxq}
    \left | \left\| \bm{b}\left(\bm{x}, \bm{p} \right) \right \| - \left\| \bm{b}\left(\bm{x}, \bm{q} \right) \right \| \right | \le \left\| \bm{b}\left(\bm{x}, \bm{p} \right) -  \bm{b}\left(\bm{x}, \bm{q} \right) \right \|.
\end{equation}

Substituting \cref{eqn:bxp_byp} to \cref{eqn:abs_bxp_byp} and \cref{eqn:bxp_bxq} to \cref{eqn:abs_bxp_bxq}, it is concluded that the norm function $ \left\| \bm{b}\left(\bm{x}, \bm{p} \right) \right \|$ is separated Lipschitz continuous with the same coefficients $L_{\bm{x}}^{\bm{b}}$ and $L_{\bm{p}}^{\bm{b}}$ given above, respectively.\\

\textit{Part 2: Lipschitz continuity of $\mathcal{J}_{\bm{b}}\left(\bm{x}, \bm{p}\right)$.}\label{part:Lipschitz_Jacobian_b}

In fact, the component $\mathcal{J}_{\bm{x}, \bm{b}}\left(\bm{x}, \bm{p}\right) = \bm{W}\bm{\mathcal{M}}\left(\bm{p} \right)$ of $\mathcal{J}_{\bm{b}}\left(\bm{x}, \bm{p}\right) $ is bounded above by $\left \|\bm{W} \right \| \cdot B_{\bm{\mathcal{M}}}$ and additionally an invariant in $\bm{x}$. Hence, this is Lipschitz continuous in $\bm{x}$ with any positive scalar $L_{\bm{x}}^{\mathcal{J}_{\bm{x}, \bm{b}}}$ is the coefficient. Also, this is Lipschitz continuous in $\bm{p}$ with the coefficient $L_{\bm{p}}^{\mathcal{J}_{\bm{x}, \bm{b}}} = \left \| \bm{W}  \right \| \cdot B_{\nabla \bm{\mathcal{M}}}$. Similarly, let us consider the component $\mathcal{J} _{\bm{p}, \bm{b}}\left(\bm{x}, \bm{p}\right) = \bm{W} \left[\nabla \bm{\mathcal{M}}\left(\bm{p} \right) \bm{x} \right]$ of $\mathcal{J}_{\bm{b}}$ that is bounded above by $\left \| \bm{W} \right \|  \cdot B_{\nabla \bm{\mathcal{M}}\bm{x}}$. This is Lipschitz continuous in $\bm{x}$ and $\bm{p}$ with the corresponding coefficients $L_{\bm{p}}^{\mathcal{J}_{\bm{x}, \bm{b}}} = \left \|\bm{W} \right \| \cdot B_{\nabla \bm{\mathcal{M}}}$ and $L_{\bm{p}}^{\mathcal{J}_{\bm{p}, \bm{b}}} = \left \|\bm{W} \right \|\cdot B_{\nabla^2 \bm{\mathcal{M}}\bm{x}}$, respectively. Consequently, the two components $\mathcal{J}_{\bm{x}, \bm{b}}\left(\bm{x}, \bm{p}\right)$ and $\mathcal{J}_{\bm{p}, \bm{b}}\left(\bm{x}, \bm{p}\right)$ of $\mathcal{J}_{\bm{b}}\left(\bm{x}, \bm{p}\right)$ are separated Lipschitz continuous in the compact extended domain $\mathcal{D} = \mathcal{X} \times \mathcal{P}$ with the two corresponding pairs of corresponding coefficients $\left \{L_{\bm{x}}^{\mathcal{J}_{\bm{x}, \bm{b}}} , L_{\bm{p}}^{\mathcal{J}_{\bm{x}, \bm{b}}} \right \}$ and  $\left \{L_{\bm{x}}^{\mathcal{J}_{\bm{p}, \bm{b}}}, L_{\bm{p}}^{\mathcal{J}_{\bm{p}, \bm{b}}} \right \} $, respectively. In conclusion, so is $\mathcal{J}_{\bm{b}} \left(\bm{x}, \bm{p} \right)$.\\

\textit{Part 3: Lipschitz continuity of $g\left(\bm{x}, \bm{p}\right)$.}\label{part:Lipschitz_g}

Now we prove that $g \left(\bm{x}, \bm{p}\right)$ is also separated Lipschitz continuous. First,
\begin{align}
     | g\left(\bm{x}, \bm{p} \right) &- g\left(\bm{y}, \bm{p} \right) | \\ \nonumber
    &= \frac{1}{2} \left |  \left \|\bm{W} {\bm{\mathcal{M}}}\left(\bm{p}\right)\bm{x} - \bm{b} \right \|^2 - \left \|\bm{W} {\bm{\mathcal{M}}}\left(\bm{p}\right)\bm{y} - \bm{b} \right \|^2 \right | \\ \nonumber
    &= \frac{1}{2} \left \{ \left \|\bm{W} {\bm{\mathcal{M}}}\left(\bm{p}\right)\bm{x} - \bm{b} \right \| + \left \|\bm{W} {\bm{\mathcal{M}}}\left(\bm{p}\right)\bm{y} - \bm{b} \right \|  \right \} \cdot \left | \left \|\bm{W} {\bm{\mathcal{M}}}\left(\bm{p}\right)\bm{x} - \bm{b} \right \| - \left \|\bm{W} {\bm{\mathcal{M}}}\left(\bm{p}\right)\bm{y} - \bm{b} \right \|    \right | \\ \nonumber
    &\le \frac{1}{2} \left \{ \left \|\bm{W} {\bm{\mathcal{M}}}\left(\bm{p}\right)\bm{x} - \bm{b} \right \| + \left \|\bm{W} {\bm{\mathcal{M}}}\left(\bm{p}\right)\bm{y} - \bm{b} \right \|  \right \} \cdot \left \|\bm{W} {\bm{\mathcal{M}}}\left(\bm{p}\right)\bm{x}  - \bm{W} {\bm{\mathcal{M}}}\left(\bm{p}\right)\bm{y} \right \|    \\ \nonumber
    &= \left \{  \left \| \bm{W} \right \| \cdot B_{{\bm{\mathcal{M}}}} \cdot B_{\text{vol}} + \left \|\bm{b} \right \| \right \} \cdot \left \| \bm{b}\left(\bm{x}, \bm{p} \right) - \bm{b}\left(\bm{y}, \bm{p} \right) \right \|.
\end{align}

Substituting \cref{eqn:bxp_byp} to this inequality, we obtain:
\begin{equation}
    | g\left(\bm{x}, \bm{p} \right) - g\left(\bm{y}, \bm{p} \right) | \le  \left \{  \left \| \bm{W} \right \| \cdot B_{{\bm{\mathcal{M}}}} \cdot B_{\text{vol}} + \left \|\bm{b} \right \| \right \} \cdot \left \|\bm{W} \right \| \cdot B_{\bm{\mathcal{M}}} \cdot \left \|\bm{x} -\bm{y} \right \|.
\end{equation}

As a result, $g\left(\bm{x}, \bm{p} \right)$ is Lipschitz continuous in $\bm{x}$ with the coefficient
\begin{equation}
    L_{\bm{x}}^{g} =  \left \| \bm{W}\right \| \cdot B_{\bm{\mathcal{M}}} \cdot  \left \{ \left \| \bm{W} \right \| \cdot B_{{\bm{\mathcal{M}}}}  \cdot B_{\text{vol}} + \left \|\bm{b} \right \| \right \}  .
\end{equation}

Second, 
\begin{align}
     | g\left(\bm{x}, \bm{p} \right) &- g\left(\bm{x}, \bm{q} \right) | \\ \nonumber
    &= \frac{1}{2} \left |  \left \|\bm{W} {\bm{\mathcal{M}}}\left(\bm{p}\right)\bm{x} - \bm{b} \right \|^2 - \left \|\bm{W} {\bm{\mathcal{M}}}\left(\bm{q}\right)\bm{x} - \bm{b} \right \|^2 \right | \\ \nonumber
    &= \frac{1}{2} \left \{ \left \|\bm{W} {\bm{\mathcal{M}}}\left(\bm{p}\right)\bm{x} - \bm{b} \right \| + \left \|\bm{W} {\bm{\mathcal{M}}}\left(\bm{q}\right)\bm{x} - \bm{b} \right \|  \right \} \left | \left \|\bm{W} {\bm{\mathcal{M}}}\left(\bm{p}\right)\bm{x} - \bm{b} \right \| - \left \|\bm{W} {\bm{\mathcal{M}}}\left(\bm{q}\right)\bm{x} - \bm{b} \right \|    \right | \\ \nonumber
    &\le  \left \{\left \| \bm{W} \right \|  \cdot B_{{\bm{\mathcal{M}}}} \cdot B_{\text{vol}} + \left \|\bm{b} \right \| \right \} \cdot  \left \|\bm{W} {\bm{\mathcal{M}}}\left(\bm{p}\right)\bm{x} - \bm{W}{\bm{\mathcal{M}}}\left(\bm{q}\right)\bm{x} \right \| \\ \nonumber
    &= \left \{\left \| \bm{W} \right \|  \cdot B_{{\bm{\mathcal{M}}}} \cdot B_{\text{vol}} + \left \|\bm{b} \right \| \right \} \cdot  \left \| \bm{b}\left(\bm{x}, \bm{p}\right) - \bm{b}\left(\bm{x}, \bm{q}\right) \right \|.
\end{align}

Substitute \cref{eqn:bxp_bxq} to this product, we obtain:
\begin{equation}
    | g\left(\bm{x}, \bm{p} \right) - g\left(\bm{x}, \bm{q} \right) | \le  \left \{  \left \| \bm{W} \right \| \cdot B_{{\bm{\mathcal{M}}}} \cdot B_{\text{vol}} + \left \|\bm{b} \right \| \right \} \cdot \left \| \bm{W}\right \| \cdot B_{\nabla {\bm{\mathcal{M}}}\bm{x}} \cdot \left \|\bm{p} -\bm{q} \right \|.
\end{equation}

Consequently, $g\left(\bm{x}, \bm{p} \right)$ is Lipschitz continuous in $\bm{p}$ with the coefficient
\begin{equation}
    L_{\bm{p}}^{\bm{g}} =  \left \| \bm{W}\right \| \cdot B_{\nabla {\bm{\mathcal{M}}}\bm{x}} \cdot  \left \{ \left \| \bm{W} \right \| \cdot B_{{\bm{\mathcal{M}}}}  \cdot B_{\text{vol}} + \left \|\bm{b} \right \| \right \}  .
\end{equation}

In conclusion, the objective function $g\left(\bm{x}, \bm{p} \right)$ is separated Lipschitz continuous in the compact extended domain $\mathcal{D} = \mathcal{X} \times \mathcal{P}$ with the corresponding coefficients $L_{\bm{x}}^{\bm{g}}$ and $L_{\bm{p}}^{\bm{g}}$.
\end{proof}

\begin{proposition}\label{prop:Lipschitz_grad_norm_b}
Assuming $\bm{x}$ is not vanished in the iterations of the proposed gradient method. Hence, so is the projection function $\bm{b}\left(\bm{x}, \bm{p} \right)$. In the other word, there exists $\delta_{\bm{b}} > 0 $ such that $\left \| \bm{b}\left( \bm{x}, \bm{p}\right) \right \| > \delta_{\bm{b}}$. In this scenario, the partial derivative $\nabla_{\bm{p}} \left\| \bm{b}\left(\bm{x}, \bm{p} \right) \right \| $ is separated Lipschitz continuous.
\end{proposition}

\begin{proof}
In fact, the condition $\bm{x} \ne \bm{0}$ does not contradicts the compactness of the extended domain ${\mathcal{X}}$. Because $\bm{b}\left(\bm{x}, \bm{p} \right) \ne \bm{0}$, $\left \| \bm{b}\left(\bm{x}, \bm{p} \right) \right \|$ is differentiable everywhere with the gradient:
\begin{equation}
    \nabla _{\bm{p}} \left \|\bm{b}\left(\bm{x}, \bm{p} \right) \right \| = \frac{\mathcal{J}_{\bm{p}, \bm{b}}\left(\bm{p}\right)^{\top} \bm{b}\left(\bm{x}, \bm{p} \right)}{\left \|\bm{b}\left(\bm{x}, \bm{p}\right) \right \|} = \mathcal{J}_{\bm{p}, \bm{b}}\left(\bm{x}, \bm{p}\right)^{\top}\bm{u}\left(\bm{x},  \bm{p}\right),
\end{equation}
where $\bm{u}\left(\bm{x}, \bm{p} \right):= \bm{b}\left(\bm{x}, \bm{p}\right) / \left\|\bm{b}\left(\bm{x}, \bm{p}\right) \right \|$ is the unit vector in the direction of $\bm{b}\left(\bm{x}, \bm{p}\right)$, and $\mathcal{J}_{\bm{p}, \bm{b}}\left(\bm{x}, \bm{p} \right)$ is the Jacobian of $\bm{b}\left(\bm{x},  \bm{p} \right)$ with respect to $\bm{p}$. \\

\textit{Part 1: Lipschitz continuity of $\nabla_{\bm{p}} \left\| \bm{b}\left(\bm{x}, \bm{p} \right) \right \|$  in $\bm{x}$.} 

Indeed, for any $\bm{p}\in \mathcal{P}$, for all $\bm{x}, \bm{y} \in \mathcal{X}^* := \mathcal{X} \setminus \left \{ \bm{0} \right \}$, we have:
\begin{align}
    &\left \| \nabla _{\bm{p}} \left \|\bm{b}\left(\bm{x}, \bm{p} \right) \right \| - \nabla _{\bm{p}} \left \|\bm{b}\left(\bm{y}, \bm{p} \right) \right \| \right \| \\ 
    & \hspace*{1cm}= \left \| \mathcal{J}_{\bm{p}, \bm{b}}\left(\bm{x}, \bm{p}\right)^{\top}\bm{u}\left( \bm{x}, \bm{p}\right) - \mathcal{J}_{\bm{p}, \bm{b}}\left(\bm{y}, \bm{p}\right)^{\top}\bm{u}\left(\bm{y},  \bm{p}\right) \right \| \\
    &\hspace*{1cm}= \left \| \left [ \mathcal{J}_{\bm{p}, \bm{b}}\left(\bm{x}, \bm{p}\right)^{\top} - \mathcal{J}_{\bm{p}, \bm{b}}\left(\bm{y}, \bm{p}\right)^{\top} \right]\bm{u}\left(\bm{x},  \bm{p}\right) + \mathcal{J}_{\bm{p}, \bm{b}}\left(\bm{y}, \bm{p}\right)^{\top} \left[ \bm{u}\left(\bm{x},  \bm{p}\right) - \bm{u}\left(\bm{y},  \bm{p}\right) \right] \right \| \\
    &\hspace*{1cm} \le \left \|\mathcal{J}_{\bm{p}, \bm{b}}\left(\bm{x}, \bm{p}\right) - \mathcal{J}_{\bm{p}, \bm{b}}\left(\bm{y}, \bm{p}\right) \right \| \cdot \left \| \bm{u}\left(\bm{x},  \bm{p}\right) \right \| + \left \| \mathcal{J}_{\bm{p}, \bm{b}}\left(\bm{y}, \bm{p}\right) \right \| \cdot \left \| \bm{u}\left(\bm{x},  \bm{p}\right) - \bm{u}\left(\bm{y},  \bm{p}\right) \right \|   \\ \label{eqn:diff_gradp_bx_by}
    &\hspace*{1cm}= \left \|\mathcal{J}_{\bm{p}, \bm{b}}\left(\bm{x}, \bm{p}\right) - \mathcal{J}_{\bm{p}, \bm{b}}\left(\bm{y}, \bm{p}\right) \right \| + \left \| \mathcal{J}_{\bm{p}, \bm{b}}\left(\bm{y}, \bm{p}\right) \right \| \cdot \left \| \bm{u}\left(\bm{x},  \bm{p}\right) - \bm{u}\left(\bm{y},  \bm{p}\right) \right \|.
\end{align}

By the \textit{Part 2} of the \cref{prop:Mpx}, the first term of the sum in the equality \cref{eqn:diff_gradp_bx_by} is Lipschitz continuous with the coefficient $L_{\bm{x}}^{\mathcal{J}_{\bm{p}, \bm{b}}}$. Moreover, $ \left \| \mathcal{J}_{\bm{p}, \bm{b}}\left(\bm{y}, \bm{p}\right) \right \| \le \left \| \bm{W} \right \| \cdot  B_{\nabla \bm{\mathcal{M}}\bm{x}}$. Now we bound the last term $\left \| \bm{u}\left(\bm{x}, \bm{p}\right) - \bm{u}\left(\bm{y},  \bm{p}\right) \right \|$:
\begin{align}
    \left \| \bm{u}\left(\bm{x}, \bm{p}\right) - \bm{u}\left(\bm{y}, \bm{p}\right) \right \| &= \left \| \frac{\bm{b}\left(\bm{x}, \bm{p}\right)}{\left\|\bm{b}\left(\bm{x}, \bm{p}\right) \right\|} - \frac{\bm{b}\left(\bm{y}, \bm{p}\right)}{\left\|\bm{b}\left(\bm{y}, \bm{p}\right) \right\|} \right \| \\
    &= \left \| \frac{\bm{b}\left(\bm{x}, \bm{p}\right) - \bm{b}\left(\bm{y}, \bm{p}\right)}{\left\|\bm{b}\left(\bm{x}, \bm{p}\right) \right\|} + \left[\frac{1}{\left \|\bm{b}\left(\bm{x}, \bm{p}\right) \right \|} - \frac{1}{\left\|\bm{b}\left(\bm{y}, \bm{p}\right) \right\|} \right] \bm{b}\left(\bm{y}, \bm{p}\right) \right \| \\
    & \le \left \| \frac{\bm{b}\left(\bm{x}, \bm{p}\right) - \bm{b}\left(\bm{y}, \bm{p}\right)}{\left\|\bm{b}\left(\bm{x}, \bm{p}\right) \right\|} \right \| + \left \| \frac{1}{\left \|\bm{b}\left(\bm{x}, \bm{p}\right) \right \|} - \frac{1}{\left\|\bm{b}\left(\bm{y}, \bm{p}\right) \right\|} \right \| \cdot \left \|  \bm{b}\left(\bm{y}, \bm{p}\right) \right \| \\ 
    & = \frac{2}{ \left \| \bm{b}\left(\bm{x}, \bm{p} \right) \right \|} \left \|\bm{b}\left(\bm{x}, \bm{p}\right) - \bm{b}\left(\bm{y}, \bm{p}\right)  \right \| \\\label{eqn:bound_uxp_uyp}
    & \le \frac{2}{\delta_{\bm{b}}} \left \|\bm{b}\left(\bm{x}, \bm{p}\right) - \bm{b}\left(\bm{y}, \bm{p}\right)  \right \|.
\end{align}

Substitute \cref{eqn:bxp_byp} to \cref{eqn:bound_uxp_uyp}, we obtain:
\begin{equation}
    \left \| \bm{u}\left(\bm{x},  \bm{p}\right) - \bm{u}\left(\bm{y},  \bm{p}\right) \right \| \le \frac{2}{\delta_{\bm{b}}} \cdot\left \|\bm{W} \right \| \cdot B_{\bm{\mathcal{M}}} \cdot \left \|\bm{x} -\bm{y} \right \|. 
\end{equation}

Consequently, the component $\nabla_{\bm{p}} \left\| \bm{b}\left(\bm{x}, \bm{p} \right) \right \| $ of the gradient function $\nabla \left\| \bm{b}\left(\bm{x}, \bm{p} \right) \right \| $ is Lipschitz continuous in $\bm{x}$ with the coefficient
\begin{equation}
L_{\bm{x}}^{\nabla_{\bm{p}}\left \| \bm{b} \right \|} = L_{\bm{x}}^{\mathcal{J}_{\bm{p}, \bm{b}}} + \frac{2}{\delta_{\bm{b}}} \cdot \left \|\bm{W} \right \|^2 \cdot B_{\bm{\mathcal{M}}} \cdot B_{\nabla \bm{\mathcal{M}}\bm{x}} .    
\end{equation} 

\textit{Part 2: Lipschitz continuity of $\nabla_{\bm{p}} \left\| \bm{b}\left(\bm{x}, \bm{p} \right) \right \|$  in $\bm{p}$.} 

Similarly, for any $\bm{x}\in \mathcal{X}^*$, for all $\bm{p}, \bm{q} \in \mathcal{P}$, it yields:
\begin{align}
    &\left \| \nabla _{\bm{p}} \left \|\bm{b}\left(\bm{x}, \bm{p} \right) \right \| - \nabla _{\bm{p}} \left \|\bm{b}\left(\bm{x}, \bm{q} \right) \right \| \right \| \\ 
    \label{eqn:diff_grads}
    &\hspace*{1cm} \le \left \|\mathcal{J}_{\bm{p}, \bm{b}}\left(\bm{x}, \bm{p}\right) - \mathcal{J}_{\bm{p}, \bm{b}}\left(\bm{x}, \bm{q}\right) \right \| + \left \| \mathcal{J}_{\bm{p}, \bm{b}}\left(\bm{x}, \bm{q}\right) \right \| \cdot \left \| \bm{u}\left(\bm{x},  \bm{p}\right) - \bm{u}\left(\bm{x},  \bm{q}\right) \right \|.
\end{align}

By the \textit{Part 2} of the \cref{prop:Mpx}, the first term of the sum in \cref{eqn:diff_grads} is Lipschitz continuous with the coefficient $L_{\bm{p}}^{\mathcal{J}_{\bm{p}, \bm{b}}} = \left \|\bm{W} \right \| \cdot B_{\nabla ^2\bm{\mathcal{M}}\bm{x}}$. Also, $ \left \| \mathcal{J}_{\bm{p}, \bm{b}}\left(\bm{x}, \bm{q}\right) \right \| \le \left \| \bm{W} \right \| \cdot B_{\nabla \bm{\mathcal{M}}\bm{x}}$. Moreover, similarly with the calculation in the \textit{Part 1}, we have the following bound for the last term $\left \| \bm{u}\left(\bm{x}, \bm{p}\right) - \bm{u}\left(\bm{x},  \bm{q}\right) \right \|$:
\begin{align}\label{eqn:bound_uxp_uxq}
    \left \| \bm{u}\left(\bm{x},  \bm{p}\right) - \bm{u}\left(\bm{x},  \bm{q}\right) \right \|
    \le \frac{2}{\delta_{\bm{b}}} \left \|\bm{b}\left(\bm{x}, \bm{p}\right) - \bm{b}\left(\bm{x}, \bm{q}\right)  \right \|.
\end{align}

Substitute \cref{eqn:bxp_bxq} to \cref{eqn:bound_uxp_uxq}, we obtain:
\begin{equation}
    \left \| \bm{u}\left(\bm{x},  \bm{p}\right) - \bm{u}\left(\bm{x},  \bm{q}\right) \right \| \le \frac{2}{\delta_{\bm{b}}} \cdot \left\|  \bm{W} \right \| \cdot  {B_{\nabla{\bm{\mathcal{M}}}\bm{x}}} \cdot \left \| \bm{p} - \bm{q} \right \|. 
\end{equation}

Consequently, the component $\nabla_{\bm{p}} \left\| \bm{b}\left(\bm{x}, \bm{p} \right) \right \| $ of the gradient function $\nabla \left\| \bm{b}\left(\bm{x}, \bm{p} \right) \right \| $ is Lipschitz continuous in $\bm{p}$ with the coefficient
\begin{equation}
L_{\bm{p}}^{\nabla_{\bm{p}}\left \| \bm{b} \right \|} = \left \|\bm{W} \right \|  \cdot B_{\nabla ^2\bm{\mathcal{M}}\bm{x}} + \frac{2}{\delta_{\bm{b}}} \cdot \left \|\bm{W} \right \|^2 \cdot B_{\nabla \bm{\mathcal{M}}\bm{x}}^{2} .    
\end{equation} 
\end{proof}

Now we have had everything to verify the most crucial theoretical property of this mathematical study:
\begin{theorem}
The gradient $\nabla g \left(\bm{x}, \bm{p}\right)$ is separated Lipschitz continuous. Consequently, if the stepsizes chosen following the Lipschitz continuity standard, the iterative schemes \eqref{eq:GDBB_x} and \eqref{eq:GDBB_p} are then Cauchy's sequences and guaranteed to converge to a stationary point of $g\left(\bm{x}, \bm{p} \right)$.
\end{theorem}

\begin{proof}
The proof is for any $\bm{p} \in \mathcal{P}$, $\bm{q} \in \mathcal{P}$, $\bm{x} \in \mathcal{X}^*$ and $\bm{y} \in \mathcal{X}^*$, and we split it into four parts.\\

\textit{Part 1: The Lipschitz continuity of $\nabla_{\bm{x}} g\left(\bm{x}, \bm{p} \right)$ in $\bm{x}$.}
We continue to repeat the steps of verifying the conditions of the separated Lipschitz continuity. First,
\begin{align}
     \left \| \nabla_{\bm{x}} g\left(\bm{x}, \bm{p} \right) - \nabla_{\bm{x}} g\left(\bm{y}, \bm{p} \right) \right\|   &= \left \| \bm{\mathcal{M}} \left(\bm{p}  \right)^{\top} \bm{W}^{\top} \left[ \bm{W}\bm{\mathcal{M}}\left(\bm{p} \right)\bm{x}- \bm{b} \right] - \bm{\mathcal{M}} \left(\bm{p}  \right)^{\top} \bm{W}^{\top} \left[ \bm{W}\bm{\mathcal{M}}\left(\bm{p} \right)\bm{y}- \bm{b} \right] \right \| \\
     &= \left \|  \bm{\mathcal{M}} \left(\bm{p}  \right)^{\top} \bm{W}^{\top}\bm{W}\bm{\mathcal{M}}\left(\bm{p} \right) \left[\bm{x} - \bm{y} \right] \right \| \\
     & \le \left \|  \bm{\mathcal{M}} \left(\bm{p}  \right)^{\top} \bm{W}^{\top}\bm{W}\bm{\mathcal{M}}\left(\bm{p} \right) \right \| \cdot  \left \| \bm{x} - \bm{y} \right \| \\ 
     &\le B_{\bm{\mathcal{M}}}^{2} \cdot \left \| \bm{W} \right \|^{2} \cdot \left\|\bm{x} - \bm{y} \right \| .
\end{align}

Hence, $\nabla _{\bm{x}}g\left(\bm{x}, \bm{p} \right)$ is Lipschitz continuous in $\bm{x}$ with the coefficient 
\begin{equation}
    {L}_{\bm{x}}^{\nabla_{\bm{x}} g} = B_{\bm{\mathcal{M}}}^2 \cdot \left \| \bm{W} \right \|^2.
\end{equation}

\textit{Part 2: The Lipschitz continuity of $\nabla_{\bm{x}} g\left(\bm{x}, \bm{p} \right)$ in $\bm{p}$.} Moreover,
\begin{align}
     \left \| \nabla_{\bm{x}} g\left(\bm{x}, \bm{p} \right) - \nabla_{\bm{x}} g\left(\bm{x}, \bm{q} \right) \right\|   &= \left \| \bm{\mathcal{M}} \left(\bm{p}  \right)^{\top} \bm{W}^{\top} \left[ \bm{W}\bm{\mathcal{M}}\left(\bm{p} \right)\bm{x}- \bm{b} \right] - \bm{\mathcal{M}} \left(\bm{q}  \right)^{\top} \bm{W}^{\top} \left[ \bm{W}\bm{\mathcal{M}}\left(\bm{q} \right)\bm{x}- \bm{b} \right] \right \| \\ \label{eqn:diff_grad_x_p_q}
     & \le \left \|\left[\bm{\mathcal{M}}\left(\bm{p} \right) - \bm{\mathcal{M}}\left(\bm{q} \right) \right]^{\top} \bm{W}^{\top} \bm{b}\right \| \\ \nonumber
     &\hspace{0.5cm}+ \left \| \left \{ \left[\bm{W}\bm{\mathcal{M}}\left(\bm{p}\right)\right]^{\top}\bm{W}\bm{\mathcal{M}}\left(\bm{p}\right) - \left[\bm{W}\bm{\mathcal{M}}\left(\bm{q}\right)\right]^{\top}\bm{W}\bm{\mathcal{M}}\left(\bm{q}\right) \right \} \bm{x} \right \| .
\end{align}

The first term of the sum in the inequality \cref{eqn:diff_grad_x_p_q} satisfies the following property:
\begin{equation}
    \left \|\left[\bm{\mathcal{M}}\left(\bm{p} \right) - \bm{\mathcal{M}}\left(\bm{q} \right) \right]^{\top} \bm{W}^{\top} \bm{b}\right \| \le B_{\nabla \bm{\mathcal{M}}} \cdot \left \| \bm{W} \right \| \cdot \left \| \bm{b} \right \| \cdot \left \|\bm{p} - \bm{q} \right \|.
\end{equation}

Put $\bm{v}\left(\bm{p}\right) = \bm{W}\bm{\mathcal{M}}\left(\bm{p}\right)$ and $\bm{v}\left(\bm{q}\right) = \bm{W}\bm{\mathcal{M}}\left(\bm{q}\right)$, then we have $\left \|\bm{v}\left(\bm{p}\right) \right \| \le \left \| \bm{W} \right \| \cdot B_{\bm{\mathcal{M}}}$, $\left \|\bm{v}\left(\bm{q}\right) \right \| \le \left \| \bm{W} \right \| \cdot B_{\bm{\mathcal{M}}}$, and $\left \|\bm{v}\left(\bm{p}\right) - \bm{v}\left(\bm{q}\right)  \right \| \le \left \| \bm{W} \right \| \cdot B_{\nabla \bm{\mathcal{M}}} \cdot \left \| \bm{p} - \bm{q} \right \| $. As a result, the second term of the sum in the inequality \cref{eqn:diff_grad_x_p_q} satisfies the following property:
\begin{align}
    & \left \| \left \{ \left[\bm{W}\bm{\mathcal{M}}\left(\bm{p}\right)\right]^{\top}\bm{W}\bm{\mathcal{M}}\left(\bm{p}\right) - \left[\bm{W}\bm{\mathcal{M}}\left(\bm{q}\right)\right]^{\top}\bm{W}\bm{\mathcal{M}}\left(\bm{q}\right) \right \} \bm{x} \right \| \\
    &\hspace*{0.5cm} = \left \| \left \{\left[ \bm{v}\left(\bm{p}\right) \right ]^{\top} \bm{v}\left(\bm{p}\right) - \left[ \bm{v}\left(\bm{q}\right) \right ]^{\top} \bm{v}\left(\bm{q}\right)\right \} \bm{x} \right \| \\
    &\hspace*{0.5cm} = \left \| \left \{\left[ \bm{v}\left(\bm{p}\right) - \bm{v}\left(\bm{q}\right) \right ]^{\top} \bm{v}\left(\bm{p}\right) + \left[ \bm{v}\left(\bm{q}\right) \right ]^{\top}  \left[\bm{v}\left(\bm{p}\right) - \bm{v}\left(\bm{q}\right) \right]\right \} \bm{x} \right \| \\
    &\hspace*{0.5cm} \le \left [  \left \|\bm{v}\left(\bm{p}\right) - \bm{v}\left(\bm{q}\right)  \right \| \cdot \left \|\bm{v}\left(\bm{p}\right) \right \| + \left \|\bm{v}\left(\bm{q}\right) \right \| \cdot \left \|\bm{v}\left(\bm{p}\right) - \bm{v}\left(\bm{q}\right)  \right \| \right ] \cdot \left \|  \bm{x} \right \|\\
    &\hspace*{0.5cm} \le \left \| \bm{W} \right \| \cdot B_{\bm{\mathcal{M}}}   \cdot B_{\text{vol}} \cdot \left \|\bm{v}\left(\bm{p}\right) - \bm{v}\left(\bm{q}\right)  \right \| \\
    &\hspace*{0.5cm} \le \left \| \bm{W} \right \|^2 \cdot B_{\bm{\mathcal{M}}} \cdot B_{\nabla \bm{\mathcal{M}}}   \cdot B_{\text{vol}} \cdot \left \| \bm{p} - \bm{q}  \right \|.
\end{align}

Hence, $\nabla _{\bm{x}}g\left(\bm{x}, \bm{p} \right)$ is Lipschitz continuous in $\bm{p}$ with the coefficient
\begin{equation}
    {L}_{\bm{p}}^{\nabla_{\bm{x}} g} = \left \| \bm{W} \right \|^2 \cdot B_{\bm{\mathcal{M}}} \cdot B_{\nabla \bm{\mathcal{M}}}   \cdot B_{\text{vol}}.
\end{equation}

\textit{Part 3: The Lipschitz continuity of $\nabla_{\bm{p}} g\left(\bm{x}, \bm{p} \right)$ in $\bm{x}$.} Similarly,
\begin{align}
     &\left \| \nabla_{\bm{p}} g\left(\bm{x}, \bm{p} \right) - \nabla_{\bm{p}} g \left(\bm{y}, \bm{p} \right) \right \| \\
     &= \left \| \left[\nabla\bm{\mathcal{M}} \left(\bm{p} \right)\bm{x} \right]^{\top} \bm{W}^{\top} \left[ \bm{W}\bm{\mathcal{M}}\left(\bm{p} \right)\bm{x}- \bm{b} \right] -  \left[\nabla\bm{\mathcal{M}} \left(\bm{p} \right)\bm{y} \right]^{\top} \bm{W}^{\top} \left[ \bm{W}\bm{\mathcal{M}}\left(\bm{p} \right)\bm{y}- \bm{b} \right]\right \|\\ \label{eqn:diff_grad_p_x_y}
     &\le  \left \|\left \{ \nabla\bm{\mathcal{M}} \left(\bm{p} \right) \left[\bm{x} - \bm{y} \right] \right \}^{\top} \bm{W}^{\top} \bm{b} \right \| \\ \nonumber
     &\hspace*{0.5cm} +\left \|\left[\nabla\bm{\mathcal{M}} \left(\bm{p} \right)\bm{x} \right]^{\top} \bm{W}^{\top}  \bm{W}\bm{\mathcal{M}}\left(\bm{p} \right)\bm{x} - \left[\nabla\bm{\mathcal{M}} \left(\bm{p} \right)\bm{y} \right]^{\top} \bm{W}^{\top} \bm{W}\bm{\mathcal{M}}\left(\bm{p} \right)\bm{y} \right \|.
\end{align}

The first term of the sum in the inequality \cref{eqn:diff_grad_p_x_y} satisfies the following property:
\begin{equation}
    \left \|\left \{ \nabla\bm{\mathcal{M}} \left(\bm{p} \right) \left[\bm{x} - \bm{y} \right] \right \}^{\top} \bm{W}^{\top} \bm{b} \right \| \le \left \|\bm{W} \right \| \cdot \left \| \bm{b} \right \| \cdot B_{\nabla \bm{\mathcal{M}}} \cdot \left \| \bm{x} - \bm{y} \right \|.
\end{equation}

Beside, the second term of the sum in the inequality \cref{eqn:diff_grad_p_x_y} satisfies the following property:
\begin{align}
    &\left \|\left[\nabla\bm{\mathcal{M}} \left(\bm{p} \right)\bm{x} \right]^{\top} \bm{W}^{\top}  \bm{W}\bm{\mathcal{M}}\left(\bm{p} \right)\bm{x} - \left[\nabla\bm{\mathcal{M}} \left(\bm{p} \right)\bm{y} \right]^{\top} \bm{W}^{\top} \bm{W}\bm{\mathcal{M}}\left(\bm{p} \right)\bm{y} \right \| \\
    & = \frac{1}{2} \left \| \nabla _{\bm{p}} \left \| \bm{b}\left(\bm{x}, \bm{p} \right) \right \|^2 - \nabla _{\bm{p}} \left \| \bm{b}\left(\bm{y}, \bm{p} \right) \right \|^2 \right \| \\
    & = \frac{1}{2} \left \| \nabla _{\bm{p}} \left[  \left \| \bm{b}\left(\bm{x}, \bm{p} \right) \right \|^2 -  \left \| \bm{b}\left(\bm{y}, \bm{p} \right) \right \|^2 \right] \right \| \\ \label{eqn:second_grad_p_x_y}
    & \le \frac{1}{2}\left[  \left \| \bm{b}\left(\bm{x}, \bm{p} \right) \right \| + \left \| \bm{b}\left(\bm{y}, \bm{p} \right) \right \| \right] \cdot \left \| \nabla _{\bm{p}} \left[  \left \| \bm{b}\left(\bm{x}, \bm{p} \right) \right \| -  \left \| \bm{b}\left(\bm{y}, \bm{p} \right) \right \| \right] \right \|  \\ \nonumber
    & \hspace*{0.5cm}+ \frac{1}{2} \left |   \left \| \bm{b}\left(\bm{x}, \bm{p} \right) \right \| - \left \| \bm{b}\left(\bm{y}, \bm{p} \right) \right \| \right | \cdot \left \| \nabla _{\bm{p}} \left[  \left \| \bm{b}\left(\bm{x}, \bm{p} \right) \right \| +  \left \| \bm{b}\left(\bm{y}, \bm{p} \right) \right \| \right] \right \|.
\end{align}

Let us recall the bound $\left \| \bm{b} \left(\cdot, \bm{p} \right) \right \| \le \left \| \bm{W} \right \| \cdot B_{\bm{\mathcal{M}}} \cdot B_{\text{vol}}$. By the {\textit{Part 1}} of the \cref{prop:Lipschitz_grad_norm_b}, the first term in the sum in the inequality \cref{eqn:second_grad_p_x_y} is then Lipschitz continuous with the coefficient $ \left \| \bm{W} \right \| \cdot B_{\bm{\mathcal{M}}} \cdot B_{\text{vol}} \cdot L_{\bm{x}}^{\nabla_{\bm{p}}\left \| \bm{b} \right \|}$. Beside, due to the boundedness of all the operators and the variables, there exists an upper bound $B_{\nabla_{\bm{p}} \left \| \bm{b} \right \|}< +\infty$ for the norm of each term  $\nabla _{\bm{p}} \left \| \bm{b}\left(\bm{x}, \bm{p} \right) \right \|$ and $\nabla_{\bm{p}} \left \| \bm{b}\left(\bm{y}, \bm{p} \right) \right \|$, i.e.,:
\begin{equation}
      \sup_{\bm{p} \in \mathcal{P}, \bm{x} \in \mathcal{X}}  \left\| {\nabla _{\bm{p}} } \left \| \bm{b} \left( \bm{x}, \bm{p} \right) \right \| \right\|  \le B_{\nabla_{\bm{p}} \left \| \bm{b} \right \|}  \text{ and}  \sup_{\bm{p} \in \mathcal{P}, \bm{y} \in \mathcal{X}}  \left\| {\nabla _{\bm{p}} } \left \| \bm{b} \left( \bm{y}, \bm{p} \right) \right \| \right\|  \le B_{\nabla_{\bm{p}} \left \| \bm{b} \right \|} .
\end{equation}

Hence, the second term in the sum in the inequality \cref{eqn:second_grad_p_x_y} is then Lipschitz continuous with the coefficient $L_{\bm{x}}^{\bm{b}} \cdot B_{\nabla_{\bm{p}} \left \| \bm{b} \right \|}$. Consequently, $\nabla _{\bm{p}}g\left(\bm{x}, \bm{p} \right)$ is Lipschitz continuous in $\bm{x}$ with the coefficient 
\begin{equation}
{L}_{\bm{x}}^{\nabla_{\bm{p}} g} = \left \|\bm{W} \right \| \cdot \left \| \bm{b} \right \| \cdot B_{\nabla \bm{\mathcal{M}}} + \left \| \bm{W} \right \| \cdot B_{\bm{\mathcal{M}}} \cdot B_{\text{vol}} \cdot L_{\bm{x}}^{\nabla_{\bm{p}}\left \| \bm{b} \right \|} + L_{\bm{x}}^{\bm{b}} \cdot B_{\nabla_{\bm{p}} \left \| \bm{b} \right \|}.    
\end{equation}

\textit{Part 4: The Lipschitz continuity of $\nabla_{\bm{p}} g\left(\bm{x}, \bm{p} \right)$ in $\bm{p}$.} Finally,
\begin{align}
     &\left \| \nabla_{\bm{p}} g\left(\bm{x}, \bm{p} \right) - \nabla_{\bm{p}} g \left(\bm{x}, \bm{q} \right) \right \| \\
     &= \left \| \left[\nabla\bm{\mathcal{M}} \left(\bm{p} \right)\bm{x} \right]^{\top} \bm{W}^{\top} \left[ \bm{W}\bm{\mathcal{M}}\left(\bm{p} \right)\bm{x}- \bm{b} \right] -  \left[\nabla\bm{\mathcal{M}} \left(\bm{q} \right)\bm{x} \right]^{\top} \bm{W}^{\top} \left[ \bm{W}\bm{\mathcal{M}}\left(\bm{q} \right)\bm{x}- \bm{b} \right]\right \|\\ \label{eqn:RHS_norms}
     & \le \left\| \left[\nabla \bm{\mathcal{M}} \left(\bm{p} \right) \bm{x} - \nabla \bm{\mathcal{M}} \left(\bm{q} \right) \bm{x} \right]^{\top}\bm{W}^{\top} \bm{b} \right\|  \\ \nonumber
     & \hspace*{0.5cm} + \left \|  \left[\nabla\bm{\mathcal{M}} \left(\bm{p} \right)\bm{x} \right]^{\top} \bm{W}^{\top}  \bm{W}\bm{\mathcal{M}}\left(\bm{p} \right)\bm{x}  -  \left[\nabla\bm{\mathcal{M}} \left(\bm{q} \right)\bm{x} \right]^{\top} \bm{W}^{\top} \bm{W}\bm{\mathcal{M}}\left(\bm{q} \right)\bm{x} \right \|.
\end{align}

Again, let us recall the intermediate value theorem in Banach spaces:
\begin{equation}
 \left \| {\nabla \bm{\mathcal{M}}} (\bm{p}) \bm{x} - {\nabla \bm{\mathcal{M}}}(\bm{q})\bm{x} \right \|  \le {B_{\nabla^2 {\bm{\mathcal{M}}}\bm{x}}} \left \| \bm{p} - \bm{q} \right \|.   
\end{equation}

Since, the first term of the sum in the inequality \cref{eqn:RHS_norms} satisfies the following property:
\begin{equation}
     \left\| \left[\nabla \bm{\mathcal{M}} \left(\bm{p} \right) \bm{x} - \nabla \bm{\mathcal{M}} \left(\bm{q} \right) \bm{x} \right]^{\top}\bm{W}^{\top} \bm{b} \right\| \le {B_{\nabla^2 {\bm{\mathcal{M}}}\bm{x}}} \cdot \left \|\bm{W} \right \| \cdot \left \| \bm{b} \right \| \cdot \left \| \bm{p} - \bm{q} \right \|.
\end{equation}

Hence, this is Lipschitz continuous with the coefficient ${B_{\nabla^2 {\bm{\mathcal{M}}}\bm{x}}} \cdot \left \|\bm{W} \right \| \cdot \left \| \bm{b} \right \|$. Now let us consider the second term of the sum in the inequality \cref{eqn:RHS_norms}:
\begin{align}
    &\left \|  \left[\nabla\bm{\mathcal{M}} \left(\bm{p} \right)\bm{x} \right]^{\top} \bm{W}^{\top}  \bm{W}\bm{\mathcal{M}}\left(\bm{p} \right)\bm{x}  -  \left[\nabla\bm{\mathcal{M}} \left(\bm{q} \right)\bm{x} \right]^{\top} \bm{W}^{\top} \bm{W}\bm{\mathcal{M}}\left(\bm{q} \right)\bm{x} \right \| \\
    &= \frac{1}{2}\left \| \nabla_{\bm{p}} \left \|\bm{W}\bm{\mathcal{{M}}}\left(\bm{p} \right) \bm{x}\right \|^2 -   \nabla_{\bm{p}} \left \|\bm{W}\bm{\mathcal{{M}}}\left(\bm{q} \right) \bm{x}\right \|^2 \right \| \\
    &= \frac{1}{2}\left \| \nabla_{\bm{p}} \left \|\bm{b}\left(\bm{x}, \bm{p} \right)\right \|^2 -   \nabla_{\bm{p}} \left \|\bm{b}\left(\bm{x}, \bm{q} \right)\right \|^2 \right \| \\
    &= \frac{1}{2} \left \| \nabla_{\bm{p}} \left \{  \left[ \left \|\bm{b}\left(\bm{x}, \bm{p} \right)\right \| + \left \|\bm{b}\left(\bm{x}, \bm{q} \right)\right \| \right] \cdot \left[\left \|\bm{b}\left(\bm{x}, \bm{p} \right)\right \| - \left \|\bm{b}\left(\bm{x}, \bm{q} \right)\right \|\right] \right\} \right \| \\
    & \le \frac{1}{2}  \left \|\left[ \left \|\bm{b}\left(\bm{x}, \bm{p} \right)\right \| + \left \|\bm{b}\left(\bm{x}, \bm{q} \right)\right \| \right] \cdot\nabla_{\bm{p}} \left[ \left \|\bm{b}\left(\bm{x}, \bm{p} \right)\right \| - \left \|\bm{b}\left(\bm{x}, \bm{q} \right)\right \| \right] \right \| \\ \nonumber
    & \hspace*{0.5cm} +\frac{1}{2}  \left \|\left[ \left \|\bm{b}\left(\bm{x}, \bm{p} \right)\right \| - \left \|\bm{b}\left(\bm{x}, \bm{q} \right)\right \| \right] \cdot\nabla_{\bm{p}} \left[ \left \|\bm{b}\left(\bm{x}, \bm{p} \right)\right \| + \left \|\bm{b}\left(\bm{x}, \bm{q} \right)\right \| \right] \right \| \\ \label{eqn:second_term}
    & \le \left \| \bm{W} \right \| \cdot B_{\bm{\mathcal{M}}} \cdot B_{\text{vol}} \cdot \left \| \nabla_{\bm{p}} \left[ \left \|\bm{b}\left(\bm{x}, \bm{p} \right)\right \| - \left \|\bm{b}\left(\bm{x}, \bm{q} \right)\right \| \right]\right \| \\ \nonumber
    & \hspace*{0.5cm} +\frac{1}{2}  \left \|\left[ \left \|\bm{b}\left(\bm{x}, \bm{p} \right)\right \| - \left \|\bm{b}\left(\bm{x}, \bm{q} \right)\right \| \right] \cdot\nabla_{\bm{p}} \left[ \left \|\bm{b}\left(\bm{x}, \bm{p} \right)\right \| + \left \|\bm{b}\left(\bm{x}, \bm{q} \right)\right \| \right] \right \|.
\end{align}
By the \textit{Part 2} of the \cref{prop:Lipschitz_grad_norm_b}, the first term of the sum in the inequality \cref{eqn:second_term} is then Lipschitz continuous with the coefficient 
\begin{equation}
    {L}_{\bm{p}, 2}^{\nabla g} = \left \| \bm{W} \right \| \cdot B_{\bm{\mathcal{M}}} \cdot B_{\text{vol}} \cdot L_{\bm{p}}^{\nabla_{\bm{p}} \left \| \bm{b} \right \|}.
\end{equation}

 Beside, the second term of the sum in the inequality \cref{eqn:second_term} satisfies the following property:
\begin{align}
&\frac{1}{2}\left \|\left[ \left \|\bm{b}\left(\bm{x}, \bm{p} \right)\right \| - \left \|\bm{b}\left(\bm{x}, \bm{q} \right)\right \| \right] \cdot\nabla_{\bm{p}} \left[ \left \|\bm{b}\left(\bm{x}, \bm{p} \right)\right \| + \left \|\bm{b}\left(\bm{x}, \bm{q} \right)\right \| \right] \right \| \\ \label{eqn:last}
&\le \frac{1}{2} \left\| \nabla_{\bm{p}} \left[ \left \|\bm{b}\left(\bm{x}, \bm{p} \right)\right \| + \left \|\bm{b}\left(\bm{x}, \bm{q} \right)\right \| \right] \right \| \cdot \left | \left[ \left \|\bm{b}\left(\bm{x}, \bm{p} \right)\right \| - \left \|\bm{b}\left(\bm{x}, \bm{q} \right)\right \| \right] \right |.
\end{align}

We use the bound $B_{\nabla \left \| \bm{b} \right \|}$ defined in the \textit{Part 3} of this theorem for the first term of the product in the inequality \cref{eqn:last}. Moreover, by the \textit{Part 1} of the \cref{prop:Mpx}, the second term of the product in the inequality \cref{eqn:last} is Lipschitz continuous in $\bm{p}$ with the coefficient $L_{\bm{p}}^{\bm{b}}$. Hence, the second term of the sum in the inequality \cref{eqn:second_term} is Lipschitz continuous with the coefficient $B_{\left \|\nabla \bm{b} \right \|} \cdot L_{\bm{p}}^{\bm{b}}$. As a result, so is the Lipschitz continuity of the second term in the inequality \cref{eqn:RHS_norms} with the coefficient ${L}_{\bm{p}, 2}^{\nabla g} + B_{\left \|\nabla \bm{b} \right \|} \cdot L_{\bm{p}}^{\bm{b}}$. Consequently, $\nabla_{\bm{p}} g\left(\bm{x}, \bm{p} \right)$ is Lipschitz continuous in $\bm{p}$ with the coefficient
\begin{equation}
    {L}_{\bm{p}}^{\nabla_{\bm{p}} g} = {B_{\nabla^2 {\bm{\mathcal{M}}}\bm{x}}} \cdot \left \|\bm{W} \right \| \cdot \left \| \bm{b} \right \| + {L}_{\bm{p}, 2}^{\nabla g} + B_{\left \|\nabla \bm{b} \right \|} \cdot L_{\bm{p}}^{\bm{b}}.
\end{equation}

In conclusion, $\nabla g\left(\bm{x}, \bm{p} \right)$ is separated Lipschitz continuous in $\bm{x}$ and $\bm{p}$ with the coefficients ${L}_{\bm{x}}^{\nabla_{\bm{x}} g}$, ${L}_{\bm{p}}^{\nabla_{\bm{x}} g}$, ${L}_{\bm{x}}^{\nabla_{\bm{p}} g}$, and ${L}_{\bm{p}}^{\nabla_{\bm{p}} g}$ computed in the above four parts. The sequences $\left \{\bm{x}^{i} \right \}_{i=1}^{+\infty}$ and $\left \{\bm{p}^{i} \right \}_{i=1}^{+\infty}$ collected from the iterative schemes \eqref{eq:GDBB_x} and \eqref{eq:GDBB_p} with the chosen stepsizes satisfying Lipschitz continuity standard are then Cauchy's sequences and guaranteed to converge to a stationary point of $g\left(\bm{x}, \bm{p} \right)$.
\end{proof}

\subsection{Initialization}
\subsubsection{Initial stepsizes}\label{subsec:stepsize}
 In practice, the fixed stepsizes satisfy the Lipschitz continuity condition do not accelerate the convergence. Alternatively, if the size of the object volumes and the number of subscans are small, for instance, around $64 \times 64\times 64$ (voxel) and 5 subscans, the stepsizes $\gamma_{\bm{x}}^k$ and $\gamma_{\bm{p}}^k$ in the first iterations can be chosen by a backtracking line search until the Armijo condition \cite{Armijo66} is fulfilled, and the line search can be executed on personal single-GPU computers \cite{Nguyen22_SPIE}. Nonetheless, line search strategies become impractical for larger-scale volumes, such as those of size approximately $500 \times 500 \times 500$ (voxel) in this study. To address this issue, the stepsizes can instead be initialized as a scaled proportion of the normalized values ${1}/{\left \|\nabla {\msquare} g \right \|}$ using suitable proportional coefficients $c^0{\msquare}$, i.e., $\gamma_{\msquare}^0 = c_{\msquare}^0/{\left \|\nabla _{\msquare} g \right \|}$. Subsequently, the Barzilai–Borwein formula \cite{BB88} is employed in the following iterations, leveraging information from the entire scan to achieve accelerated convergence compared with line search strategies \cite{Hansen21}, as applied in the schemes \cref{eq:GDBB_x} and \cref{eq:GDBB_p}:
\begin{equation}\label{eq:BB}
    \gamma_{\msquare}^k = \frac{\langle \nabla _{\msquare} g\left(\bm{x}^k, \bm{p}^k\right) -\nabla _{\msquare} g\left(\bm{x}^{k-1}, \bm{p}^{k-1}\right) , {{\Msquare}}^k - {{\Msquare}}^{k-1} \rangle}{\left \| \nabla _{\msquare} g\left(\bm{x}^k, \bm{p}^k\right) -\nabla _{\msquare} g\left(\bm{x}^{k-1}, \bm{p}^{k-1}\right) \right \|^2},
\end{equation}
where ${\Box} = \bm{x}$ or $\bm{p}$. 

It is important to note that this formula is used for calculating the stepsizes in all our SAMCIRT experiments. Also, it is defined since the second iteration and it requires an appropriate two-point value of both $\bm{x}$ and $\bm{p}$ obtained after the initialization and the first optimization step, before being applied to iterate for the stepsize values of the following iterations. Also, the objective function does not always decrease due to the instability of the Barzilai-Borwein method.

\subsubsection{Initial reconstruction}
A trivial guess for the motion parameter $\bm{p}$ is the value corresponding to the static cases of the global motion operator $\bm{\mathcal{M}}$. As a result, the derivative $\nabla_{\bm{p}} g \left(\bm{x}, \bm{p}^0 \right)$ is vanished. However, the first optimization step is flexibly done so that there is still a small change of $\bm{p}$ in the first iteration. Since the motion domain $\mathcal{P}$ is compact, the intermediate estimates of $\bm{p}$ do not need to be back-projected onto the same domain after each iteration. We call the \textit{reconstruction without motion correction} as the reconstruction obtained by executing the scheme \cref{eq:GDBB_x} when this static motion is taken into account. This initial guess is useful for SAMCIRT if the motion is small and the image is not very blurry. This is also considered with respect to optimization in terms of minimization. In particular, it corresponds to the solution of the following minimization problem:
\begin{equation}
    \bm{x}^0 = \text{argmin}_{\bm{x}} \left \| \bm{W} \bm{x} - \bm{b} \right \|^2,
\end{equation}
which can be solved by any least-square solver (e.g., LSQR, BFGS, or GMBB (gradient method with the stepsizes computed by the Barzilai-Borwein formula)). This initial image is expected not very blurry as the motion is small and there has not been any evidence that it influences the convergence of affine transformation.

\subsection{Optimization criteria}
The following metric is used to evaluate the quality of the 4DCT reconstruction:
\begin{definition}
Let $\widehat{\bm{b}}_i$ and $\bm{b}_i$ be the re-projection with estimated motion parameters beforehand and the projection data in the real acquisition, respectively. The re-projection and the projection here can be corresponding to one or more projection angles depending on the choice of the subscan. The \textbf{projection distance} (PD) is the quantity defined by the $l_2$-norm of the \textbf{projection difference}:
\begin{equation}
    PD = \left \|\widehat{\bm{b}}_i - \bm{b}_i \right \|.
\end{equation}
    
In particular, for SAMCIRT the PD is given as follows:
\begin{equation}
    PD = \left \|\bm{W}_i \mathcal{M}\left(\widehat{\bm{p}}_i\right) \widehat{\bm{x}} - \bm{b}_i \right \|,
\end{equation}    
where $\widehat{\bm{x}}$ and $\widehat{\bm{p}}_i$ are the simultaneous reconstruction and the motion parameters corresponding to the $i^{th}$ projection, respectively.

The metric quantifies the difference between the acquired projection data and the projection computed using the motion-compensated reconstruction. In other words, it evaluates how well the final reconstruction and the estimated motion parameters satisfy projection data fidelity.
\end{definition}

\section{Datasets}\label{sec:datasets}
This section details the datasets employed to evaluate the proposed method. The experimental data was acquired with the TESCAN UniTOM XL system at the Antwerp Center for 4D Quantitative X-ray Imaging and Analysis\footnote{\url{https://www.uantwerpen.be/en/research-facilities/center-for-4d-quantitative-x-ray/}}. The projection geometry in all real scans is cone beam. The flat field correction and log normalization were all done by the FlexRayTools \cite{DeSamber21}. We use the standard abbreviations SOD and SDD to denote the distance from the X-ray source to the object and to the detector, respectively. 

\begin{table}[h!]
\centering
\begin{tabular}{|c|c|c|c|}
\hline
parameter & real diamond & simulated diamond & real sponge \\
\hline
voxel size ($\mu$m) & 8  & 0.25 & 92.46 \\
projection size (pixel) & 1896 $\times$ 1920 & 310 $\times$ 320 & 1896 $\times$ 1920 \\
binning number & 4 & 1 & 4\\
volume size (voxel) & 474 $\times$ 480 $\times$ 480 & 300 $\times$ 310 $\times$ 320 & 474 $\times$ 480 $\times$ 480\\
SOD (mm) & 40 & 1000 & 462 \\
SDD (mm) & 750 & 1250 & 750 \\
acquisition time &  1h 26m 35s & not given & 5 minutes\\
motion type & continuous rigid motion & discrete rigid motion & discrete scaling \\
\hline
\end{tabular}
\caption{\centering Detail of the datasets and experiments testing SAMCIRT.}
\end{table}

\begin{figure}
    \centering
    \begin{subfigure}{0.32\linewidth}
        \includegraphics[width=\linewidth]{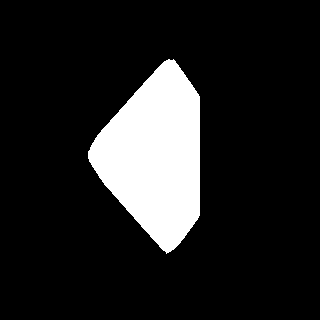}
    \end{subfigure}
    \begin{subfigure}{0.32\linewidth}
        \includegraphics[width=\linewidth]{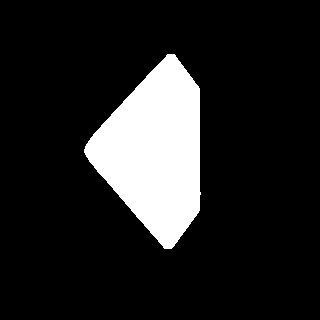}
    \end{subfigure}
    \begin{subfigure}{0.32\linewidth}
        \includegraphics[width=\linewidth]{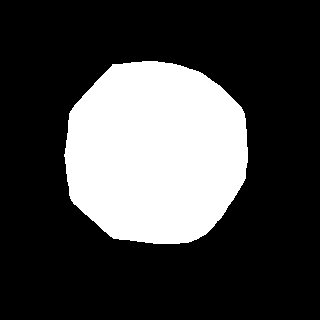}
    \end{subfigure}
    \caption{\centering central cross-sections of the 3D homogeneous diamond phantom.}
    \label{fig:binary_Diamond}
\end{figure}

\begin{figure*}[t]
    \centering
    \includegraphics[width=\linewidth]{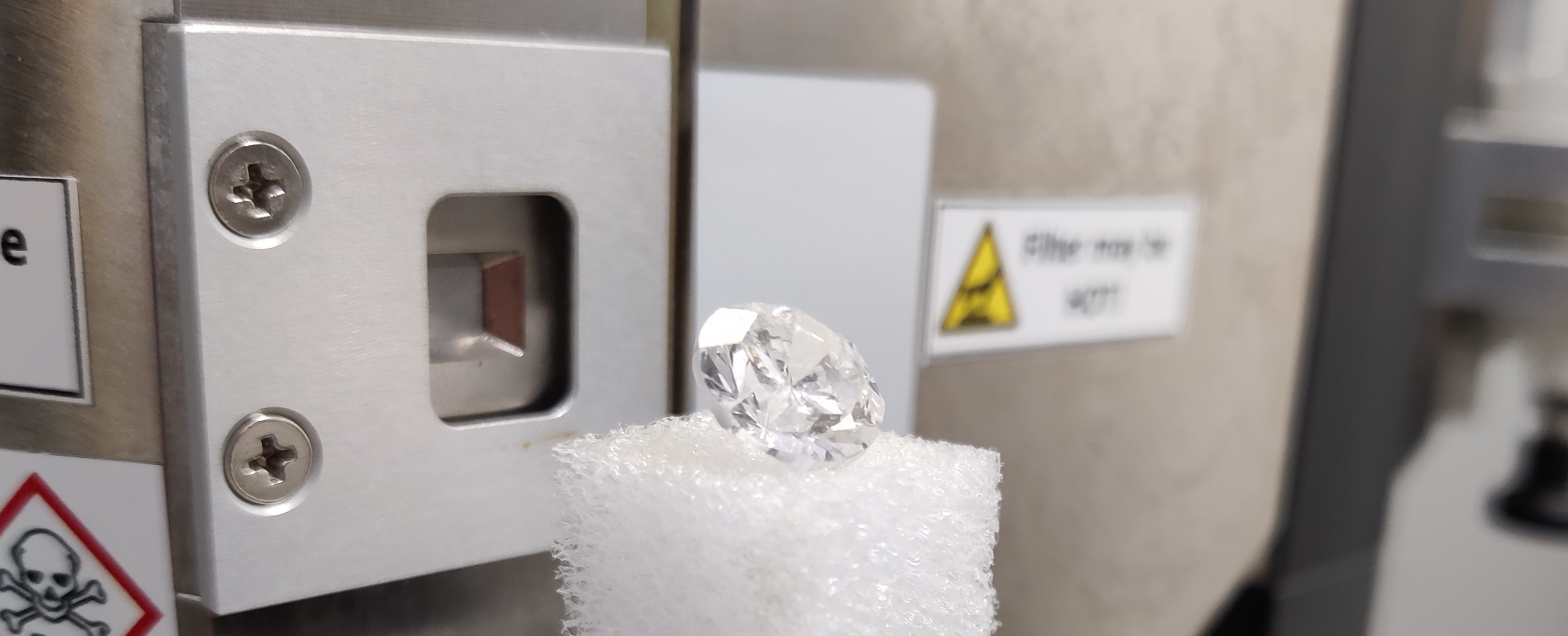}
    \caption{\centering the real diamond in front of the X-ray source of the CT scanner.}
    \label{fig:real_Diamond}
\end{figure*}

The diameter of the real round diamond (\cref{fig:real_Diamond}) is measured geometrically from the projection data and the projection geometry is 12.376 mm (around 7.3 carats). On the other hand, a homogeneous diamond phantom obtained by voxelizing an approximate surface mesh of the reconstruction without motion correction of the diamond from the given real projection dataset. The surface mesh was created using the \textit{Volume to Mesh} modifier in Blender \cite{Blender18}, followed by a remeshing operation. A visualization of the binary diamond phantom is presented in \cref{fig:binary_Diamond}. Finally, the sponge dataset consists of projections acquired from two consecutive step-and-shoot full angular-range scans. The sponge was static in the first full scan, and vertically compressed in the second full scan. The 3D scaling factor was $\widehat{\bm{s}} = \left[s_x, s_y, s_z\right]^T = \left[0.97,0.93, 1.21\right]^T$, which was estimated by solving the following problem by ImWIP \cite{Renders23ImWIP}:
\begin{equation}
    \widehat{s} = \text{argmin}_{\bm{s}\in \mathbb{R}^3} \left \| \mathcal{M}(\bm{s} )\bm{x}_1 - \bm{x}_2\right \|_2^2,
\end{equation}
where $\bm{x}_1$ and $\bm{x}_2$ are the reconstruction of the sponge from the projection data acquired in the first and the second full scans, respectively. In fact, the reconstruction $\bm{x}_1$ can be used as a ground truth to validate any 4DCT reconstruction on this dataset. The projection angles in the first and second scans are $[0, \pi)$ and $[\pi, 2\pi)$, respectively.

\section{Experiments}\label{sec:validation}
In this section, we test our proposed method on the datasets described in \cref{sec:datasets}. The SAMCIRT and the Cryo-EM reconstruction methods \cite{Zehni20} were tested on the same computer with the following hardware configuration: Intel® Core™ i7, RAM 64 GB, GPU GeForce RTX 2080 SUPER running on Linux OS. \cite{VanNieuwenhove17_TIP} was tested on a computer with similar hardware configuration but with a GeForce RTX 1070 GPU and Windows OS. Subsequently, we compared the results obtained by our method with the results obtained by the relevant affine-motion estimation methods \cite{VanNieuwenhove17_TIP, Zehni20}. Because only affine motion parameters corresponding to the deformations perpendicular to the projection direction are accurately estimated, we do not compare the estimated motion parameters with the ground truth parameters in the simulation experiments.

\subsection{Rigid motion estimation}
Our proposed method is tested on the acquisition of solid objects, including simulated projection data of the homogeneous diamond phantom and real projection data of the real diamond, respectively. 

\subsubsection{On the homogeneous diamond phantom: simulation experiment 1}\label{sec:simul_exp_1}

\begin{figure}
    \centering
    \begin{subfigure}{0.32\linewidth}
        \includegraphics[width=\linewidth]{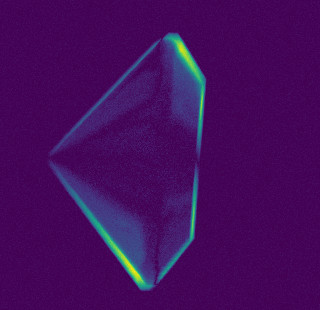}
        \caption{\centering with \cite{VanNieuwenhove17_TIP}: $PD = 3107.95$. \newline}
    \end{subfigure}
    \begin{subfigure}{0.32\linewidth}
        \includegraphics[width=\linewidth]{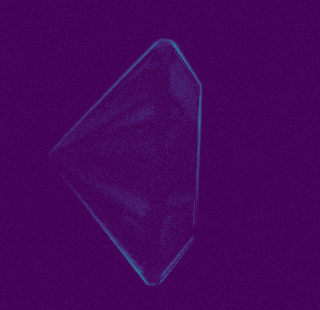}        
        \caption{\centering with SAMCIRT: $PD = 944.77$.}
    \end{subfigure}
    \caption{\centering projection difference and projection distance (PD) results on the first projection of the homogeneous diamond phantom.}
    \label{fig:vs_Vincent_binaryDiamond}
\end{figure}

This subsection provides a consistent experiment setup so that our method can be compared with \cite{VanNieuwenhove17_TIP}. We concentrate on the acceleration of the motion estimation, so only a few numbers of projections are required in both of our methods. In particular, 8 projections were used in the reference scan for the method \cite{VanNieuwenhove17_TIP}, followed by a projection chosen in the main scan in which the object of the size $300 \times 310 \times 320$ (voxel) was translated in three dimensions with the corresponding displacement $[-5, 7, 10]$ (voxel), and was rotated with the corresponding angles $[0.02, -0.05, 0.1]$ (rad). Gaussian noise with a standard deviation of $1\%$ of the gray peak value of the projection data was added to the sinogram. As only two subscans were used with a small number of projection angles, which results in a much lower-dimensional optimization problem, SAMCIRT quickly converged after 10 iterations with a running speed less than 4 sec./iter. An example of the projection differences is presented in the \cref{fig:vs_Vincent_binaryDiamond}, in which SAMCIRT outperforms \cite{VanNieuwenhove17_TIP} in projection distance. 

\subsubsection{On the homogeneous diamond phantom: simulation experiment 2}\label{sec:simul_exp_2}
This subsection establishes a consistent experimental setup to enable comparison with the method in \cite{Zehni20}, which formulates the reconstruction problem using an ADMM-based optimization scheme with a non-smooth regularization term on $\bm{x}$ and an objective function different from ours. This distinction has a significant impact on algorithmic design and computational performance. In contrast, our method (SAMCIRT) employs a smooth objective with analytically computed gradients and avoids nested iterations, providing a more direct and scalable approach. To validate this, we consider continuous motion estimation on a homogeneous diamond volume of size $320 \times 320 \times 320$ (voxel) that was acquired in a full-angular scan that contained 10 equilateral projection angles, each GD step update implemented by \cite{Zehni20} took more than 10 minutes and the same time for 200 iterations of the ADMM optimizer executed in the same method, when each iteration of SAMCIRT with the parallel beam geometry (consistently with the Cryo-EM reconstruction technique \cite{Zehni20}) took around 35 seconds by running on the same computer. 

\subsubsection{Experiment on the real diamond dataset}

\begin{figure}
    \centering
    \includegraphics[width=0.68\linewidth]{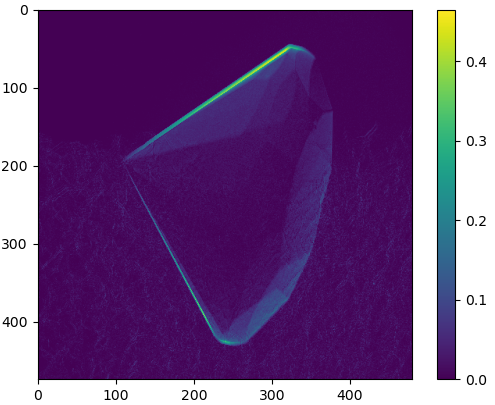}
    \caption{\centering difference of the projections at the  projection angle $0$ and the projection angle $2 \pi$ the real diamond.}
    \label{fig:diff_projs_Diamond}
\end{figure}

\begin{figure}
    \centering
    \begin{subfigure}{0.68\linewidth}
    \includegraphics[width=\linewidth]{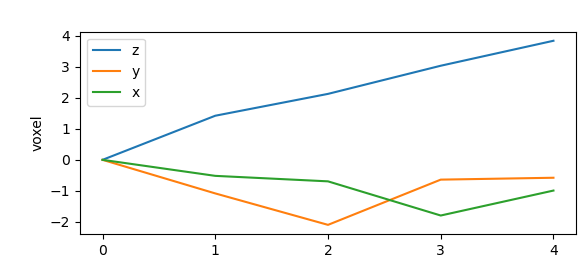}
    \end{subfigure}
    \begin{subfigure}{.68\linewidth}
    \includegraphics[width=\linewidth]{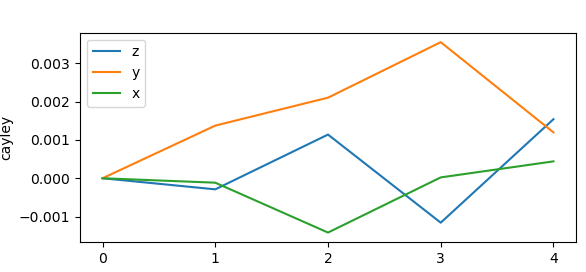}
    \end{subfigure}
    \caption{\centering parameter estimation result of the translation (top) and the rotation (bottom) on the real diamond dataset (horizontal axis: $i^{th}$ subscan).}
    \label{fig:estimation_Diamond}
\end{figure}

\begin{figure}
    \centering
    \begin{subfigure}{0.48\linewidth}
        \includegraphics[width=\linewidth]{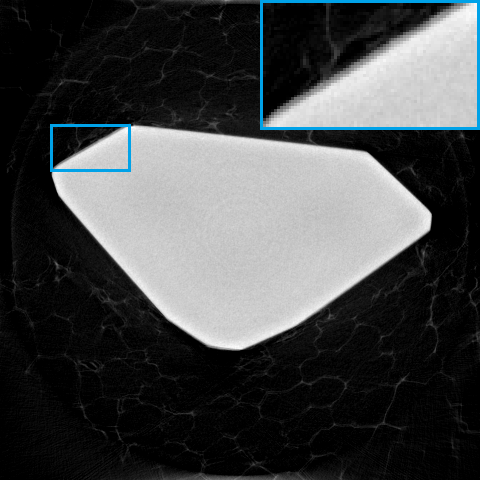}
    \end{subfigure}
    \begin{subfigure}{0.48\linewidth}
        \includegraphics[width=\linewidth]{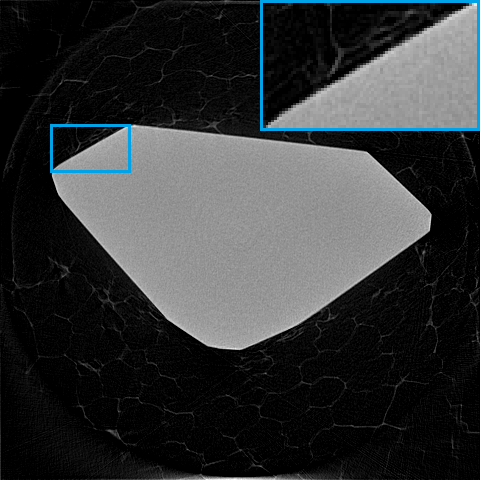}
    \end{subfigure}
    \begin{subfigure}{0.48\linewidth}
        \includegraphics[width=\linewidth]{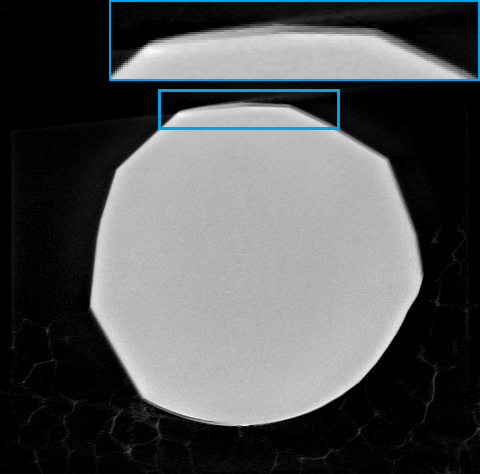}
    \end{subfigure}
    \begin{subfigure}{0.48\linewidth}
        \includegraphics[width=\linewidth]{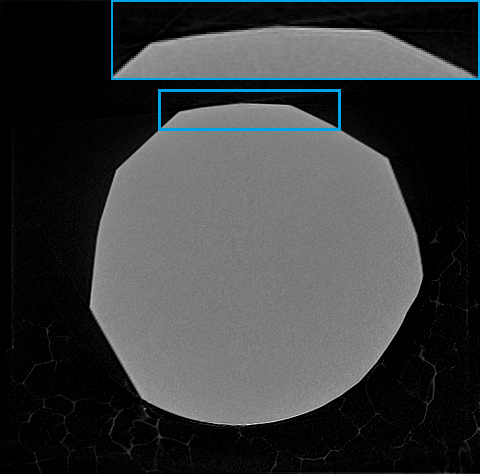}
    \end{subfigure}
    \begin{subfigure}{0.48\linewidth}
        \includegraphics[width=\linewidth]{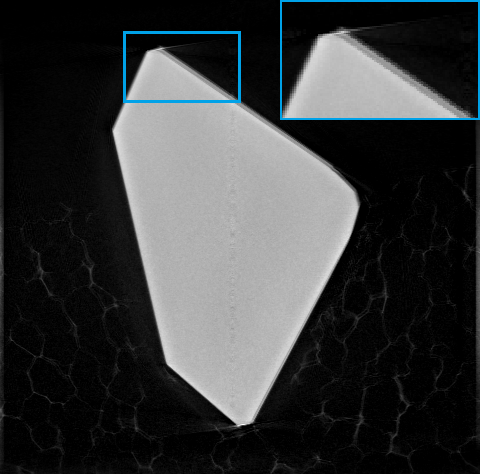}
        \caption{\centering GMBB}
    \end{subfigure}
    \begin{subfigure}{0.48\linewidth}
        \includegraphics[width=\linewidth]{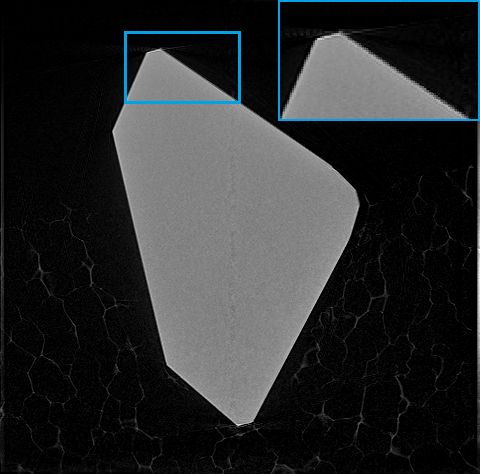}
        \caption{\centering SAMCIRT}
    \end{subfigure}
    \caption{\centering central cross-sections of the reconstruction results on the real diamond dataset.}
    \label{fig:recs_realDiamond}
\end{figure}

The binning number was set to 4, reducing the projection size to $472 \times 480$ pixels, sufficiently large for motion artifacts to remain visually discernible in reconstructions without motion correction. The object volume size was set to $472 \times 480 \times 480$ (voxel).

In conventional CT scans, the objects are usually glued with the foam put onto the scanner holder. However, for diamonds, gluing is not an option as these precious samples must be cleaned and relaxed. So, having a method in which motion is not such a problem is helpful. In our experiment, the scan duration of the diamond was long, neither the stability of the diamond onto the foam nor the foam onto the sample holder are guaranteed. In order to verify the existence of motion, let us consider the difference between the first and the last projection images. The visualization of this difference in \cref{fig:diff_projs_Diamond} shows that the diamond was not stationary during the scan. To estimate this movement by using SAMCIRT, 90 iterations were used with the proportional coefficients for the initial stepsizes of the reconstruction, rotation and translation estimation schemes are $c_{\bm{x}}^{0} = 1$, $c_{\theta}^{0} = 0.001$ and $c_{\bm{t}}^0 = 0.1$, respectively. The computation speed was recorded to be around 60 (sec./iter.). On the other hand, the ADMM optimizer in \cite{Zehni20} requires the computation or inversion of the Hessian associated with the motion operator, which introduces substantial memory demands for 3D volumes. In our implementation attempt, the method from \cite{Zehni20} was unable to run on this object volume size.

The reconstruction with compensation of rigid motion artefacts of the real diamond are shown in \cref{fig:recs_realDiamond}, in which the misalignment in the reconstruction without motion correction are corrected in the reconstruction results obtained by SAMCIRT. A remark can be made from the visualization of the difference between the first and the last projections in \cref{fig:diff_projs_Diamond} is that there could be a monotone translation in the object's vertical axis. The validation of SAMCIRT returned the estimated translation along the z-axis presented in \cref{fig:estimation_Diamond} confirmed this hypothesis.

\subsection{Scaling motion estimation}
\begin{figure}
\centering
    \begin{subfigure}{0.33\linewidth}
        \includegraphics[width=\linewidth]{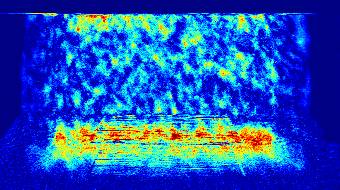}
        \caption{\centering without scaling correction: \newline $PD = 28.84$.}
    \end{subfigure}
    \begin{subfigure}{0.33\linewidth}
        \includegraphics[width=\linewidth]{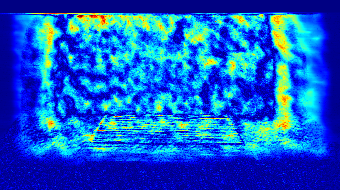}
        \caption{\centering with general affine correction  \cite{VanNieuwenhove17_TIP}: $PD = 25.63$.}
    \end{subfigure}
    \begin{subfigure}{0.33\linewidth}
        \includegraphics[width=\linewidth]{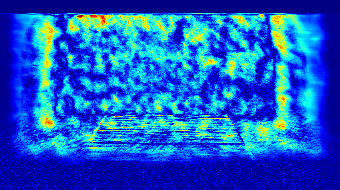}
        \caption{\centering with scaling-restricted correction  \cite{VanNieuwenhove17_TIP}: $PD = 25.33$.}
    \end{subfigure}
    \begin{subfigure}{0.33\linewidth}
        \includegraphics[width=\linewidth]{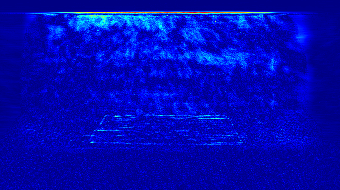}
        \caption{\centering with SAMCIRT: $PD = 9.68$. \newline}
    \end{subfigure}
    \caption{\centering projection difference and projection distance (PD) results on the sponge's real projection data.}
\label{fig:diff_vs_Vincent_sponge}
\end{figure}

Our proposed method is tested on the real sponge dataset. To validate the scaling estimation so that a consistent quantitative comparison with \cite{VanNieuwenhove17_TIP} can be made, 16 angular-ranged projections of the sponge acquired in the first full real scan and a projection acquired in the second full real scan, were used to be the data of the first and second subscans of the dynamic scan. The object volume size was set is $474 \times 480 \times 480$ (voxel) and the results on the projection difference were cropped close to the object area for better visualization. The stopping criterium of \cite{VanNieuwenhove17_TIP} was used according to the default setting of the method's authors. Similar with the simulation experiment in the \cref{sec:simul_exp_1}, the convergence is quickly achieved after only 10 iterations of SAMCIRT. Results on the projection distance, shown in \cref{fig:diff_vs_Vincent_sponge} indicates that SAMCIRT outperforms \cite{VanNieuwenhove17_TIP}.

\section{Discussion}\label{sec:discussion}
The results from both simulations and real experiments clearly demonstrate the improvement of our method compared with GMBB, which does not account for motion estimation. On the other hand, our method is validated on several real datasets when most of the state-of-the-art methods \cite{Chen19, Chung17, VanNieuwenhove17_TIP, Lucka18, Zang19, Zehni20} have not attempted yet. Furthermore, the methods \cite{VanNieuwenhove17_TIP, Lucka18} are tested on real datasets, but nested iterations complicate the computation. In particular, our method outperforms the two relevant CT reconstruction and affine motion correction techniques \cite{VanNieuwenhove17_TIP, Zehni20} in computational feasibility and \cite{VanNieuwenhove17_TIP} in projection distance.

As we initialized the motion to be identity, i.e., $\bm{p}^0 \equiv \bm{0}$ for rigid motion or $\bm{p}^0 \equiv \bm{1}$ for scaling, the partial derivative $\nabla _{\bm{p}} g \left(\bm{x}, \bm{p}^0 \right)$ is vanished no matter the reconstruction. However, we flexibly selected a sufficiently small value for this initial partial derivative in the first iteration so that there is a small change in the first iteration for estimating $\bm{p}$, before applying the two-point Barzilai-Borwein formula for iterating the stepsize values in the next iterations. As a result, in the first iterations, the component $\bm{p}_1$ of $\bm{p}$ does not significantly change from $\bm{p}_1^0$, but the reconstruction $\bm{x}$ does from the initial guess $\bm{x}^0$.

Every methods have pros and cons, and our method is not an exceptional case. While we acknowledge the complexities of real CT physics, including scatter and beam hardening, a detailed treatment of these effects is beyond the scope of this work as our proposed CT model follows standard practice as a simplified approximation to facilitate tractable analysis and algorithm development. On the other hand, in the general cases, it remains challenging to investigate the rate of convergence of the iterative schemes \cref{eq:GDBB_x} and \cref{eq:GDBB_p} and to propose a deterministic stopping criterion, because these may not be convergent to the global minima. In our experiments, we ran as many iterations as possible and monitored the process until convergence was achieved, rather than setting a specific stopping criterion. Moreover, the affine motions of non-zeros DVFs, e.g., rotation of a uniform disk at the center, cannot be estimated by the present version of our method. On the other hand, as a deliberate design choice, we evaluated whether our proposed optimization scheme could still produce accurate reconstructions and motion estimates without explicitly regularizing the reconstruction $\bm{x}$. This is motivated by our aim to assess the performance of the affine motion compensation mechanism itself. That said, in settings with higher noise levels or more complex deformation models, additional regularization on both $\bm{x}$ and $\bm{p}$ could be beneficial.

\section{Conclusion}\label{sec:conclusion}
We have presented a mathematical method for 4DCT that allows accurate reconstruction and affine motion correction. In terms of the shown results, our approach emphasizes practical applicability in areas where current state-of-the-art methods fall short. Specifically, our method surpasses existing 4DCT reconstruction and affine motion correction techniques in terms of computational feasibility and residual errors. In theoretical aspects, the findings include a mathematical foundation for supporting the convergence of the iterations proposed in our method. These are consistent among experiments on both synthetic and real datasets, and highlight a new application for 4DCT for higher resolving 3DCT reconstructions, as the state-of-the-art would allow. Another of our main contributions is that this outputs an accurate reconstruction with the rigid motion artifacts compensated for a real diamond in which SAMCIRT clearly gives better reconstructions in case of motion, which directly results in a novel application of 4DCT.

\section{Future work}\label{sec:futurework}
We have been working on developing an optimal design for selecting subscans and their corresponding projection angles. Furthermore, the joint Lipschitz continuity, a stronger condition than separated Lipschitz continuity, is being examined in the context of the objective function and its associated mappings. Also, we plan to conduct a more in-depth investigation of the convergence rate in the context of continuous motion estimation, where the solution may be global. We anticipate that this could lead to the establishment of a standard stopping criterion for the proposed iterative method, which we intend to present in a future manuscript.

\section*{Declaration on awareness of conflicts of interest}
The authors declare that neither the editor nor the reviewers suggested for this submission have had any active collaboration with the authors within the five years prior to the submission.

\section*{Acknowledgments}
This work was partially funded by the Research Foundation–Flanders (FWO), grant nos. S007219N and 1SA2920N. The authors thank Mona Zehni for making publicly available the source codes implementing the Cryo-EM reconstruction method; Cuong Tran and Tuyen Trung Truong for fruitful discussions on the mathematical foundation of the convergence proof and on improvements to the method, respectively; and Jan Sijbers and Jan De Beenhouwer for their support during different stages of this work.

\bibliographystyle{SIAMplain}
\bibliography{references}

\end{document}